\documentclass{ejm}

\usepackage{color,amsmath,graphicx,textcomp,bm,multirow,soul}
\definecolor{mike}{rgb}{1,1,0.5}
\sethlcolor{white}

\let\realverbatim=\verbatim
\let\realendverbatim=\endverbatim
\renewcommand\verbatim{\par\addvspace{6pt plus 2pt minus 1pt}\realverbatim}
\renewcommand\endverbatim{\realendverbatim\addvspace{6pt plus 2pt minus 1pt}}

\ifprodtf \else
  \checkfont{eurm10}
  \iffontfound
    \IfFileExists{upmath.sty}
      {\typeout{^^JFound AMS Euler Roman fonts on the system,
                   using the 'upmath' package.^^J}%
       \usepackage{upmath}}
      {\typeout{^^JFound AMS Euler Roman fonts on the system, but you
                   don't seem to have the}%
       \typeout{'upmath' package installed. EJM.cls can take advantage
                 of these fonts,^^Jif you use 'upmath' package.^^J}%
       \providecommand\mathup[1]{##1}%
       \providecommand\mathbup[1]{##1}%
      }
  \else
    \providecommand\mathup[1]{#1}%
    \providecommand\mathbup[1]{#1}%
  \fi
\fi

\ifprodtf \else
  \checkfont{msam10}
  \iffontfound
    \IfFileExists{amssymb.sty}
      {\typeout{^^JFound AMS Symbol fonts on the system, using the
                'amssymb' package.^^J}%
       \usepackage{amssymb}%
         \let\leq=\leqslant
         
      }{}
  \fi
\fi

\ifprodtf \else
  \IfFileExists{amsbsy.sty}
    {\typeout{^^JFound the 'amsbsy' package on the system, using it.^^J}%
     \usepackage{amsbsy}}
    {\providecommand\boldsymbol[1]{\mbox{\boldmath $##1$}}}
\fi

\newsavebox{\astrutbox}
\sbox{\astrutbox}{\rule[-5pt]{0pt}{20pt}}

\newcommand{\pd}[2]{\frac{\partial #1}{\partial #2}}

\newcommand{\oz}{^{(0)}}
\newcommand{\oo}{^{(1)}}

\newcommand{\se}[1]{\sigma_{ #1}}

\newcommand{\eff}{^{\mathrm{eff}}}

\newdefinition{definition}[theorem]{Definition}

\title[European Journal of Applied Mathematics]{Multiscale modelling and homogensation of fibre-reinforced hydrogels for tissue engineering}

\author[M. J. Chen et al.]{%
M. J. CHEN$\,^{1*}$,\ns  
L. S. KIMPTON$\,^{1*}$,\ns
  J. P. WHITELEY$\,^2$,\ns
  M. CASTILHO$\,^3$,\ns\\
 J. MALDA$\,^{3,4}$,\ns
  C. P. PLEASE$\,^1$,\ns
S. L. WATERS$\,^1$\ns
\and
  H. M. BYRNE$\,^1$
}

\affiliation{%
  $^1\,$Mathematical Institute, University of Oxford, Andrew Wiles Building,\\ Radcliffe Observatory Quarter, Woodstock Road, Oxford  OX2 6GG, UK\\
    email\textup{\nocorr: \texttt{helen.byrne@maths.ox.ac.uk}}\\
  $^2\,$Department of Computer Science, University of Oxford, Wolfson Building, Parks Road,\\
       Oxford OX1 3QD, UK\\
  $^3\,$Department of Orthopaedics, University Medical Center Utrecht, Utrecht University, Utrecht, The Netherlands\\
 $^4\,$Department of Equine Sciences, Faculty of Veterinary Medicine, Utrecht University, Utrecht, The Netherlands\\
$^*$Joint first authors}

\date{\today}
\pubyear{2000}
\volume{000}
\pagerange{\pageref{firstpage}--\pageref{lastpage}}

\begin{document}

\label{firstpage}
\maketitle

\begin{abstract}
Tissue engineering aims to grow artificial tissues \emph{in vitro} to replace those in the body that have been damaged through age, trauma or disease. A recent approach to engineer artificial cartilage involves seeding cells within a scaffold consisting of an interconnected 3D-printed lattice of polymer fibres combined with a cast or printed hydrogel, and subjecting the construct (cell-seeded scaffold) to an applied load in a bioreactor. A key question is to understand how the applied load is distributed throughout the construct. To address this, we employ homogenisation theory to derive equations governing the effective macroscale material properties of a periodic, elastic-poroelastic composite. We treat the fibres as a linear elastic material and the hydrogel as a poroelastic material, and exploit the disparate length scales (small inter-fibre spacing compared with construct dimensions) to derive macroscale equations governing the response of the composite to an applied load.  This homogenised description  reflects the orthotropic nature of the composite. To validate the model, solutions from finite element simulations of the macroscale, homogenised equations are compared to experimental data describing the unconfined compression of the fibre-reinforced hydrogels. The model is used to derive the bulk mechanical properties of a cylindrical construct of the composite material for a range of fibre spacings, and to determine the local mechanical environment experienced by cells embedded within the construct.

\end{abstract}


\section{Introduction}
\label{intro}
Tissue engineering is a rapidly developing field where one of the main goals is to generate artificial biological tissues \textit{in vitro} (for example cartilage, bone or blood vessels) \cite{Groll2016}. These tissues may then be implanted to replace natural tissues that have degenerated, been damaged, or removed during surgery. A particularly active area of this field is the development of articular cartilage implants as mature cartilage tissue has limited intrinsic capacity to heal. Cartilage damage can occur through injury or diseases such as osteoarthritis, and in the United Kingdom a third of people aged $45$ or older have sought treatment for osteoarthritis \cite{OA}. Implants must be biocompatible with native cartilage, and also able to withstand the mechanically demanding environment of a loaded joint. 

A promising direction in cartilage tissue engineering \cite{HydroZONES} involves seeding cells (mesenchymal stem cells and/or chondrocytes) on a scaffold consisting of an interconnected, 3D-printed lattice of polymer fibres combined with a cast or printed hydrogel; the seeded scaffold is then cultured in a bioreactor with biochemical and mechanical stimulation.
Reinforced hydrogel composites are an ideal material for this purpose, since they are biocompatible with cartilage cells and the elastic fibres of the lattice endow the scaffold with greater structural integrity than a scaffold made only of hydrogel \cite{Visser2015}.
The principle challenge in this approach lies in developing practical strategies that generate artificial cartilage that mimics the form and function of the natural tissue. Mathematical modelling is a valuable tool for quickly and robustly assessing the efficacy of various combinations of cell seeding strategies, biochemical and mechanical stimuli. The models can thereby guide experimental design; this is of value since these experiments are expensive, time-consuming and cannot easily be sampled at multiple time points. An important modelling question is to predict the mechanical environment and stress distribution throughout the scaffold as a first step in developing appropriate strategies to seed the scaffold with mechanosensitive cells.

\begin{figure}
     \centering
     \includegraphics[width=\textwidth]{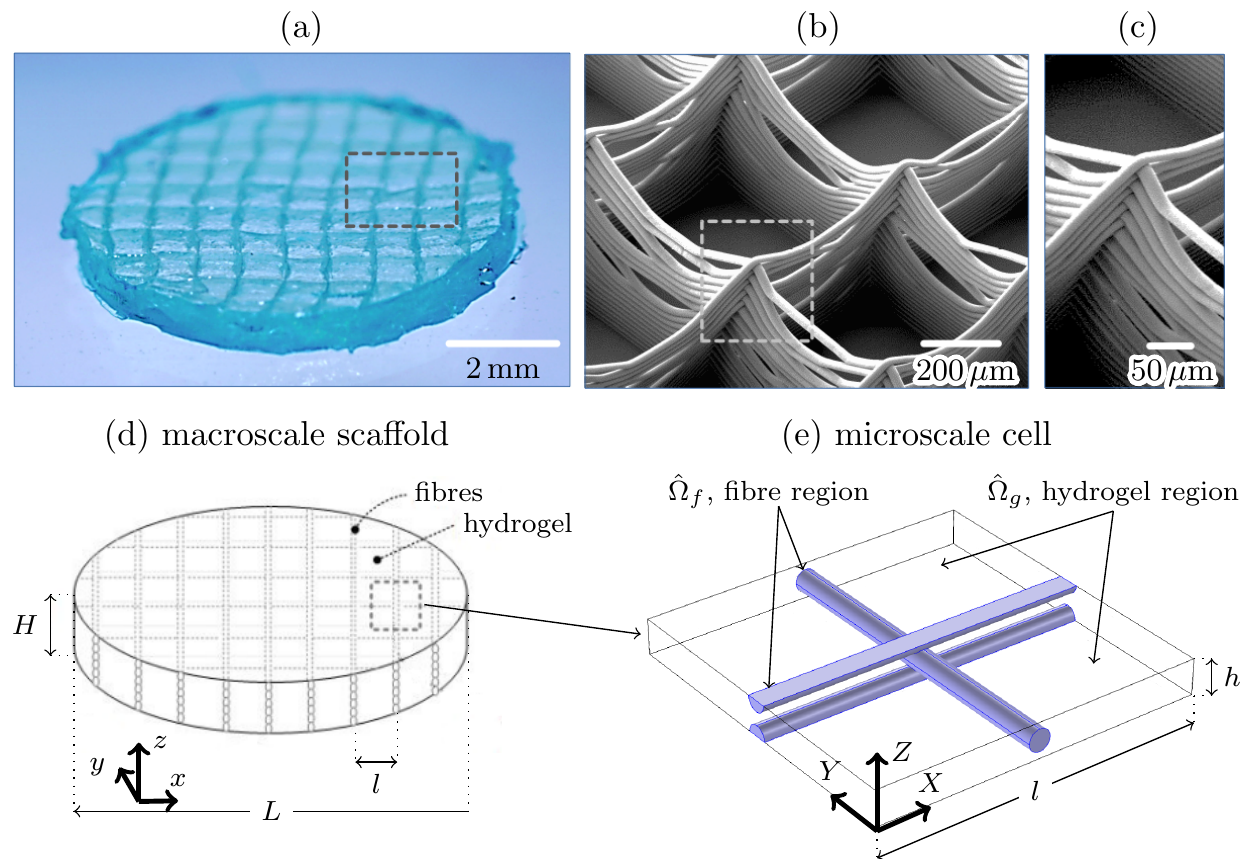}
     \caption{(a) Optical microscope image of a fibre-reinforced hydrogel with a square fibre lattice of 800\,$\mu$m. Note that the overall dimensions of the construct shown here are slightly different to those used in later experimental comparison. (b) Scanning electron microscopy (SEM) image of the fibre scaffold prior to it being cast in the hydrogel. (c) SEM image showing a detail of fibre buildup at the interconnection between printed vertical layers. (d) Schematic diagram of the idealised scaffold used in the homogenised model of this paper. (e) Schematic diagram of the microscale repeating cell, showing the microscale hydrogel region $\hat{\Omega}_g$, and the microscale fibre region $\hat{\Omega}_f$. The characteristic length scale at the microscale is the horizontal fibre spacing $l$, and the characteristic macroscale length is the overall diameter of the the scaffold $L$. It is assumed that the scaffold diameter is much greater than the fibre spacing, and that their ratio $\varepsilon=l/L\ll 1$, which permits a separation of length scales as described in Section~\ref{sec:sep-length-scales}.}\label{fig:scaffold}
 \end{figure}

The scaffold of interest in this work comprises a soft gelatin methacrylate (GelMA) hydrogel cast around a 3D-printed, $\varepsilon$-polycapralactone (PCL) fibre lattice, for details see \cite{Castilho2017,Visser2015}. The fibre lattice is created by melt electrospinning writing (MEW); a layer of parallel fibres at constant spacing is printed and then the next layer of parallel fibres at constant spacing is printed on top of the first layer, so that fibres in neighbouring layers meet at $90^{\circ}$, see Figure \ref{fig:scaffold}. The vertical distance between fibres is set by the extent to which each layer of fibres melts into the previous layer. When tested in unconfined compression, these fibre-reinforced scaffolds were shown to be up to be $54$ times stiffer (that is have a 54-fold increase in Young's modulus) than the hydrogel alone  \cite{Visser2015}. The cells that are ultimately seeded within the construct are mechanosensitive and will therefore undergo phenotypic changes due to the local stress \cite{Li2010, Sunkara2016}. Consequently, in order to understand the response of these cells to mechanical loading, it is first necessary to understand the stress induced within the fibre-reinforced hydrogel.

The fibre-reinforced hydrogel scaffold described above is an example of a composite material, combining constituent materials with known characteristics to create a new material with properties desirous for a certain application. Composite materials are prevalent in engineering, and becoming more widespread in biological applications \cite{Dunlop2015,Gladman2016,Wegst2015}. 
A natural approach to model composite materials is via mathematical homogenisation \cite{Howell2009}, which allows the macroscale response to mechanical loading of a composite material to be determined from the properties of its constituent materials and knowledge of the microstructure.

In the context of modelling the composite material of this paper, mathematical homogenisation involves writing down governing equations for the constituent materials and then exploiting the separation of length scales to decompose the full model into macroscale and periodic microscale components. This, in turn, allows the bulk effective material properties at the macroscale to be derived from the solution to a periodic microscale `cell' problem. Having determined the effective macroscale properties of the material it is possible to predict, for instance, the response of the composite material to an applied mechanical load (which is the focus of this paper). A general introduction to homogenisation theory for composite materials can be found in \cite{Howell2009}, which systematically describes approaches for treating materials with periodic microstructure for one, two and three dimensional problems. Formal asymptotic and volume averaging approaches to treating the cell problem are compared in \cite{Davit2013}. 

Homogenisation is a particularly useful tool in biological contexts, where small scale structures and multiple spatial scales are ubiquitous. In such conditions it allows tissue-level models to be derived that include cell-level properties. For example, in \cite{Shipley2010} effective transport coefficients were determined for the delivery of drugs in tumours by homogenising the microscale flow in the small scale blood vessels within the tumour. A similar approach was used to define criteria for the design of cartilage tissue engineering scaffolds in \cite{Shipley2009} by tuning the microscale properties of the scaffold to optimise the flow of nutrients. This is different to the homogenisation procedure of this paper since the goal here is to determine bulk effective mechanical properties of the scaffold.

 An alternate approach to modelling fibre-reinforced hydrogels might involve adapting an existing multiphase model of cartilage; see \mbox{\cite{Klika2016}} for a comprehensive review of such models. Fibre-reinforced hydrogels have similar mechanical properties to cartilage \mbox{\cite{Visser2015}}, so it might be argued that we should employ an existing multiphase model. However, the advantage of our homogenisation approach is that it explicitly incorporates the mechanical role of the printed fibres, and directly relates the properties of the constituent materials to those of the composite material. This then facilitates the tunable design of scaffolds with the properties required via alterations in the number, spacing and properties of the fibres.

A recent study on reinforced hydrogel composites with application to cardiac tissue engineering demonstrated that MEW can reproducibly generate fibre lattices, and that when cast in hydrogel the resulting scaffolds are biocompatible with cardiac progenitor cells \cite{Castilho2017}. Another recent study focused on the mechanical characterisation of fibre-reinforced hydrogel scaffolds, measuring the properties of both the overall scaffold and individual PCL fibres; this is of great interest since knowledge of both is required to parameterise the homogenised model of this paper. While finite element modelling of fibre-reinforced hydrogel scaffolds has previously been used to predict their overall mechanical properties \cite{Bas2017}, the homogenisation approach adopted here is more computationally efficient since it obviates the need to model each individual, repeating cell of the printed fibre lattice and the hydrogel contained within.

As stated above, we aim to understand how an applied load is distributed throughout a fibre-reinforced hydrogel construct to the embedded, mechanosensitive cells. We previously investigated the mechanics of the composite scaffold with a phenomenological model that described the stiffness of the composite \cite{Visser2015}. This simple model considered the fibres as stretched, linearly elastic strings, and neglected any rate-dependent features of the material. 

Here, we develop a more detailed model that yields greater understanding of the mechanical properties of the composite, including its time-dependent response to loading.
By developing governing equations for the stress and deformation of the composite, we develop a framework that may be used to predict the stresses that cells embedded in the scaffold experience. The resulting framework is sufficiently general that it could be adapted to predict the macroscale properties of periodic elastic-poroelastic composites in other applications.


\subsection{Paper outline}
We formulate a model for the composite material in Section \ref{sec:form}, where the fibres are treated as a linear elastic material, and the hydrogel is treated as a poroelastic material. This permits a separation of length scales, since the size of the repeating fibre lattice is much smaller than the size of the overall scaffold.  The associated microscale cell problem is described in Section \ref{sec:cell-problem}. Homogenisation theory is employed in Section \ref{sec:macroscale} to derive macroscale equations which feature effective material parameters determined from the solution to the microscale cell problem, thus determining the nature of the bulk material. This model is validated in Section \ref{sec:results-etc}, where numerical solutions of the homogenised equations are compared to unconfined compression tests on reinforced hydrogels. We discuss our results in Section \ref{sec:discuss}, where we also suggest possible future directions to continue this work.

\section{Scaffold description and model derivation}
\label{sec:form}
We aim to model the response of a fibre-reinforced hydrogel scaffold to an applied load or displacement, as discussed in \S\ref{intro}, and shown schematically in Fig.~\ref{fig:scaffold}. These scaffolds are typically a few millimetres in height and a comparable dimension in width; our model will later be compared to experimental results where cylindrical scaffolds of height $H\approx2$\,mm and diameter $L\approx5.5$\,mm are held at a strain of 6\%, for instance. Interest lies in the stress and displacement fields induced in this composite material when mechanically loaded. 

The material properties of the fibre-reinforced hydrogel, and hence its response to an applied load, will depend on the material properties of the unreinforced hydrogel, as well as the diameter and spacing of the 3D-printed fibres. These diameters and spacings are typically much smaller than the size of the overall construct; for instance, in the experiments of \cite{Visser2015} the fibres are of radius of 20\,$\mu$m and printed at fibre spacings between 200\,$\mu$m and 1\,mm. The vertical fibre spacing is difficult to determine since there is an unknown degree of melting between adjacent printed layers. In later simulations we estimate that melting results in significant overlap between the layers so that the gap between parallel fibres is 60\% of the fibre radius.

The following section details a homogenisation procedure to derive effective macroscale material properties of the reinforced construct, allowing us to calculate the stress and displacement within this composite material due to an applied load. We begin by developing sub-models for the two constituents of the composite viewing the hydrogel as a poroelastic material, occupying a region denoted $\Omega_g$, and the PCL fibres as linearly elastic, occupying a region denoted $\Omega_f$. The difference between the overall size of the construct and the spacing between the fibres permits a separation of length-scales. We exploit this property together with the periodicity of the geometry of the fibre scaffold to homogenise over one `cell' of the scaffold (see Fig.~\ref{fig:scaffold}) and obtain the desired description of this composite material.

\subsection{Sub-models for the hydrogel and the elastic fibres}

\begin{table}\begin{small}
\def\arraystretch{1.2}
\begin{tabular}{cll}
Quantity & Description & Representative value \\
\hline$\phi$ & porosity (GelMA) & (later eliminated from model) \\
$k'/\mu'$ & effective permeability (GelMA) & $2.382\times10^{-4}$kPa$^{-1}$min$^{-1}$ (Appendix \ref{sec:AppendixA}) \\
$\mu_g'$ & Lam\'{e}'s first parameter (GelMA) & 19.97\,kPa (Appendix \ref{sec:AppendixA})\\
$\lambda_g'$ & Lam\'{e}'s second parameter (GelMA) & 17.01\,kPa (Appendix \ref{sec:AppendixA})\\
$\mu_f'$ & Lam\'{e}'s first parameter (PCL) & $1.27\times10^5$\,kPa \cite{Castilho2017a}\\
$\lambda_f'$ & Lam\'{e}'s second parameter (PCL) & $7.80\times10^5$\,kPa \cite{Castilho2017a}\\
$L$ & overall diameter of scaffold & 5.54--5.98\,mm \\
$H$ & overall height of scaffold & 1.80--2.04\,mm \\
$d$ & fibre diameter & 20\,$\mu$m \\
$l$ & horizontal fibre spacing & 300--800\,$\mu$m\\
$h$ & vertical fibre spacing & 32\,$\mu$m \\
$\epsilon=l/L$ & small parameter & $5.0\times10^{-2}$--$1.4\times10^{-1}$\\
$T$ & typical test time  & 1\,min\\
$P$ & typical stress in hydrogel & $1.67\times10^{4}$\,kPa
\end{tabular}
\end{small}
\caption{Summary of dimensional parameters that appear in Equations \eqref{eq:pe-mass-dim}--\eqref{eq:kbc-dim}, along with the parameters used in the non-dimensionalisation procedure in Section \ref{sec:non-dim}.}\label{table:params}
\end{table}

Following Detournay and Cheng \cite{Detournay1993}, we describe the hydrogel as a poroelastic material comprised of incompressible fluid and elastic phases. In the hydrogel region $\Omega_g$  we have conservation of mass, and assume that the flow of the fluid phase is governed by Darcy's law. Thus, we write
 \begin{align}
\phi\bm\nabla\cdot\mathbf{v}' + (1-\phi)\pd{}{t}\left(\bm\nabla\cdot\mathbf{u}_g'\right)&=0,\label{eq:pe-mass-dim}\\
 \phi\left(\mathbf{v}'-\pd{\mathbf{u}_g'}{t'} \right)&=-\frac{k'}{\mu'}\bm\nabla p',\label{eq:pe-darcy-dim}
\end{align}
where $\mathbf{u}'_g$ is the displacement of the solid phase, $\mathbf{v}'$ is the velocity of the fluid phase and $p'$ is the fluid pressure. Equations \eqref{eq:pe-mass-dim} and \eqref{eq:pe-darcy-dim} contain several (constant) parameters, namely the volume fraction of the fluid phase, $\phi$ (sometimes called the porosity), the intrinsic permeability of the solid phase, $k'$, and the viscosity of the fluid phase, $\mu'$; the ratio of these last two parameters, $k'/\mu'$, represents the effective permeability of the poroelastic material. Typical values for these parameters for the hydrogel of interest, GelMA, are given in Table~\ref{table:params} where these were obtained by fitting data from experimental relaxation tests on unreinforced GelMA  to a model of a poroelastic material. A full description of this fitting procedure is given in Appendix \ref{sec:AppendixA}. We also require conservation of momentum in the hydrogel, and introduce a constitutive relationship between the displacement and the stress. Following \cite{Howell2009} these relationships are represented by
\begin{align}
 \bm{\nabla}\cdot\bm{\sigma}_g'&=\bm0,\label{eq:pe-mtm-dim}\\
 \bm{\sigma}_g'&=- p' \mathbf{I}+\mathbb{D}':\bm\nabla\mathbf{u}_g',\label{eq:lin-elastic-D1-dim}\\
\mathbb{D}':\bm\nabla\mathbf{u}_g'& = \mu_g'\left(\bm\nabla\mathbf{u}_g' + \left(\bm\nabla\mathbf{u}_g'\right)^T\right)+\lambda_g'\left(\bm\nabla\cdot\mathbf{u}_g'\right)\mathbf{I},\label{eq:lin-elastic-D2-dim}
 \end{align}
where $\bm{\sigma}_g'$ is the stress tensor (rank 2) in the hydrogel and $\mathbb{D}'$ is the elasticity tensor (rank 4) for the solid phase of the hydrogel. \hl{Throughout this paper we follow the conventions for tensor products and derivatives given in \mbox{\cite[Chapter 1]{Holzapfel2000}}, which also defines these conventions in Einstein notation}.
In the constitutive relationship \eqref{eq:lin-elastic-D1-dim}--\eqref{eq:lin-elastic-D2-dim} we assume that the solid phase is linearly elastic, where $\mu_g'$ and $\lambda_g'$ are the bulk Lam\'{e} parameters of the poroelastic material (which are both assumed to be constant). The fitted values of these parameters for GelMA derived in Appendix \ref{sec:AppendixA} are given in Table~\ref{table:params}; the corresponding values for the Young's modulus $E'_g$ and Poisson's ratio $\nu_g$ of the elastic phase of the hydrogel, which relate to the Lam\'{e} parameters in the standard way, are also given in Appendix \mbox{\ref{sec:AppendixA}.}

We model the PCL fibres as a linear elastic material. It is therefore straightforward to relate the stress and displacement in the fibre region $\Omega_f$ by requiring conservation of momentum and introducing an appropriate constitutive law. Following \cite{Howell2009}, for instance, we assume
\begin{align}
\nabla\cdot\bm{\sigma}_f'&=\bm0,\label{eq:fibre-mtm-dim}\\
 \bm{\sigma}_f'&=\mathbb{C}':\bm\nabla\mathbf{u}_f',\label{eq:lin-elastic-C1-dim}\\
\mathbb{C}':\bm\nabla\mathbf{u}_f'& = \mu_f'\left(\bm\nabla\mathbf{u}_f' + \left(\bm\nabla\mathbf{u}_f'\right)^T\right)+\lambda_f'\left(\bm\nabla\cdot\mathbf{u}_f'\right)\mathbf{I},\label{eq:lin-elastic-C2-dim}
\end{align}
where $\bm{\sigma}_f'$ is the stress tensor (rank 2) in the fibres, $\mathbf{u}_f'$ is the displacement in the fibre region and $\mathbb{C}'$ is the elasticity tensor (rank 4). In the constitutive relationship \eqref{eq:lin-elastic-C1-dim}--\eqref{eq:lin-elastic-C2-dim} $\mu_f'$ and $\lambda_f'$ are the (constant) Lam\'{e} parameters of this material. The values for PCL in Table~\ref{table:params} are taken from \cite{Castilho2017a}, and converted from the Young's modulus $E'_f$ and Poisson's ratio $\nu_f$ given in that study to Lam\'{e} parameters via Equation \eqref{lamedefn}.

We further assume that the fibres are perfectly bonded to the hydrogel, so that there are no voids between the fibre and gel regions. On the interface between the fibre and gel regions (denoted $\partial\Omega_f=\partial\Omega_g$) we impose continuity of stress and displacement, as well as a kinematic condition on the fluid velocity. These boundary conditions are 
\begin{align}
 \bm\sigma_g'\cdot\mathbf{n}&=\bm\sigma_f'\cdot\mathbf{n},\label{eq:cont-stress-bc-dim}\\
 \mathbf{u}_g'&=\mathbf{u}_f',\label{eq:cont-disp-bc-dim}\\
\left(\mathbf{v}'-\pd{\mathbf{u}_g'}{t'}\right)\cdot\mathbf{n}&=\bm0,\label{eq:kbc-dim}
\end{align}
on $\partial\Omega_f=\partial\Omega_g$, where $\mathbf{n}$ is the outward pointing unit normal vector to $\Omega_f$.

To summarise, the equations governing the constituent parts of this composite material consist of \eqref{eq:pe-mass-dim}--\eqref{eq:lin-elastic-D2-dim} to be solved in the poroelastic hydrogel region $\Omega_g$, and \eqref{eq:fibre-mtm-dim}--\eqref{eq:lin-elastic-C2-dim} to be solved in the elastic PCL fibre region $\Omega_f$, subject to the boundary conditions  \eqref{eq:cont-stress-bc-dim}--\eqref{eq:kbc-dim} on the interface between these regions $\partial\Omega_f=\partial\Omega_g$.

\subsection{Non-dimensionalisation}\label{sec:non-dim}
We define $L$ to be the typical diameter of a sample of the fibre-reinforced composite and $l$ to be the horizontal spacing between the printed fibres. In situations of practical interest the fibre spacing is small compared to the overall size of the composite and so we introduce the small parameter $\varepsilon$ as
\begin{align}
 \varepsilon&=\frac{l}{L}\ll 1.
\end{align}
We nondimensionalise equations \eqref{eq:pe-mass-dim}--\eqref{eq:kbc-dim}, scaling lengths with the typical diameter of the fibre-reinforced scaffold, $L$, time with a typical time scale for mechanical testing the composite, $T$, and stresses with a typical pressure in the fluid phase of the hydrogel, $P=\mu'L^2/(k'T)$. The dimensional variables (indicated by dashes) are replaced by dimensionless versions as follows
\begin{alignat}{7}
  \mathbf{u}_g'&=L\mathbf{u}_g, &\quad \mathbf{u}_f' &=L\mathbf{u}_f,&\quad p'&=Pp, &\quad t'&=Tt, \nonumber\\
\bm\sigma_g'&=P\bm\sigma_g, &\quad \bm\sigma_f'&=P\bm\sigma_f, &\quad \mathbf{x}'&=L\mathbf{x}, &\quad \mathbf{v}'&=(L/T)\mathbf{v},
\end{alignat}
and the dimensional parameters are rescaled as follows
\begin{align}
\mathbb{D}'&=P\mathbb{D}, \quad \mu_g'=P\mu_g, \quad \lambda_g'=P\lambda_g,\\ 
\mathbb{C}'&=P\mathbb{C}, \quad \mu_f'=P\mu_f, \quad \lambda_f'=P\lambda_f.
\end{align}
Under these scalings the dimensionless version of equation \eqref{eq:pe-mass-dim}--\eqref{eq:pe-darcy-dim}, which represent conservation of mass and Darcy's law in the hydrogel region $\Omega_g$, are
 \begin{align}
\phi\bm\nabla\cdot\mathbf{v} + (1-\phi)\pd{}{t}\left(\bm\nabla\cdot\mathbf{u}_g\right)&=0,\label{eq:pe-mass}\\
 \phi\left(\mathbf{v}-\pd{\mathbf{u}_g}{t} \right)&=-\bm{\nabla}p,
\end{align}
while equations \eqref{eq:pe-mtm-dim}--\eqref{eq:lin-elastic-D2-dim}, which govern conservation of momentum and the constitutive relationship, transform to give \eqref{eq:pe-mtm-dim}--\eqref{eq:lin-elastic-D2-dim} are
\begin{align}
 \bm\nabla\cdot\bm{\sigma}_g&=\bm0,\\
 \bm{\sigma}_g&=- p \mathbf{I}+\mathbb{D}:\bm\nabla\mathbf{u}_g,\label{eq:lin-elastic-D1}\\
\mathbb{D}:\nabla\mathbf{u}_g& = \mu_g\left(\bm\nabla\mathbf{u}_g + \left(\bm\nabla\mathbf{u}_g\right)^T\right)+\lambda_g\left(\bm\nabla\cdot\mathbf{u}_g\right)\mathbf{I}.\label{eq:lin-elastic-D2}
 \end{align}
In the elastic fibre region $\Omega_f$ the dimensionless versions of conservation of momentum and the constitutive relationship \eqref{eq:fibre-mtm-dim}--\eqref{eq:lin-elastic-C2-dim} are
\begin{align}
\bm\nabla\cdot\bm{\sigma}_f&=\bm0,\label{eq:fibre-mtm}\\
 \bm{\sigma}_f&=\mathbb{C}:\bm\nabla\mathbf{u}_f,\label{eq:lin-elastic-C1}\\
\mathbb{C}:\bm\nabla\mathbf{u}_f& = \mu_f\left(\bm\nabla\mathbf{u}_f + \left(\bm\nabla\mathbf{u}_f\right)^T\right)+\lambda_f\left(\bm\nabla\cdot\mathbf{u}_f\right)\mathbf{I}.\label{eq:lin-elastic-C2}
\end{align}
Finally boundary conditions \eqref{eq:cont-stress-bc-dim}--\eqref{eq:kbc-dim} transform to give
\begin{align}
 \bm\sigma_g\cdot\mathbf{n}&=\bm\sigma_f\cdot\mathbf{n},\label{eq:cont-stress-bc}\\
 \mathbf{u}_g&=\mathbf{u}_f,\\
\left(\mathbf{v}-\pd{\mathbf{u}_g}{t}\right)\cdot\mathbf{n}&=\bm0,\label{eq:kbc}
\end{align}
on $\partial\Omega_f=\partial\Omega_g$.

\subsection{Separation of length scales}\label{sec:sep-length-scales}
Having established the dimensionless governing equations and boundary conditions \eqref{eq:pe-mass}--\eqref{eq:kbc} we could, given sufficient computing resources, solve these equations numerically in the complex interpenetrating geometry defined by $\Omega_f$ and $\Omega_g$. Instead we exploit the periodic geometry and the small size of the repeating `cell' compared to that of the composite (i.e.\ $0<\varepsilon\ll 1$).
After non-dimensionalisation, typical lengths of the composite scaffold are $\mathbf{x}=O(1)$; we henceforth term this the macroscale variable. We introduce the microscale variable $\mathbf{X}=\mathbf{x}/\varepsilon$, so that $\mathbf{X}=O(1)$ is the length scale associated with the repeating cell.
Following \cite{Penta2014}, we consider that all dependent variables are functions of $\mathbf{x}$ and $\mathbf{X}$, so that e.g.\ $\boldsymbol{\sigma}_g=\boldsymbol{\sigma}_g(\mathbf{x},\mathbf{X},t)$, and treat $\mathbf{X}$ and $\mathbf{x}$ as independent variables, in which case $\nabla\rightarrow\nabla_x + \frac{1}{\varepsilon}\nabla_X$.
We also introduce regular perturbation series expansions in $\varepsilon$ for each dependent variable, so that \ $\boldsymbol{\se{}}=\boldsymbol{\se{}}\oz + \varepsilon\boldsymbol{\se{}}\oo + \mathcal{O}\left(\varepsilon^2\right)$ and so on.
Under these assumptions \eqref{eq:pe-mass}--\eqref{eq:lin-elastic-D2} supply the following leading order equations in the hydrogel region $\Omega_g$
 \begin{align}
(1-\phi)\pd{}{t}\left(\bm\nabla_X\cdot\mathbf{u}_g^{(0)}\right)+\phi\left(\bm\nabla_X\cdot\mathbf{v}^{(0)}\right) &=0,\label{eq:pe-mass-0}\\
\bm\nabla_Xp^{(0)}&=\bm0,\quad\implies p^{(0)}\equiv p^{(0)}(\mathbf{x},t),\label{eq:pe-darcy-0}\\
 \bm\nabla_X\cdot\bm\sigma^{(0)}_g&=\bm0,\label{eq:pe-mtm-0}\\
 \mathbb{D}:\bm\nabla_X\mathbf{u}_g^{(0)}&=\bm0.\label{eq:lin-elastic-D1-0}
 \end{align}
In the fibre region $\Omega_f$, equations \eqref{eq:fibre-mtm}--\eqref{eq:lin-elastic-C1} supply
\begin{align}
\bm\nabla_X\cdot\bm{\sigma}^{(0)}_f&=\bm0,\label{eq:fibre-mtm-0}\\
\mathbb{C}:\bm\nabla_X\mathbf{u}^{(0)}_f&=\bm0,\label{eq:lin-elastic-C1-0}
\end{align}
while on  $\partial\Omega_f$ boundary conditions \eqref{eq:cont-stress-bc}--\eqref{eq:kbc} supply at leading-order
\begin{align}
 \bm\sigma_g^{(0)}\cdot\mathbf{n}&= \bm\sigma_f^{(0)}\cdot\mathbf{n},\label{eq:cont-stress-bc-0}\\
 \mathbf{u}_g^{(0)} &=\mathbf{u}_f^{(0)},\label{eq:cont-disp-bc-0}\\
 \left(\mathbf{v}^{(0)}-\pd{\mathbf{u}_g^{(0)}}{t}\right)\cdot\mathbf{n}&=0.\label{eq:kbc-0}
\end{align}
Similarly, in the hydrogel region $\Omega_g$, at $\mathcal{O}(\varepsilon)$ equations \eqref{eq:pe-mass}--\eqref{eq:lin-elastic-D2} supply
 \begin{align}
 (1-\phi)\pd{}{t}\left(\bm\nabla_x\cdot\mathbf{u}_g^{(0)}\right) + \phi\bm\nabla_x\cdot\mathbf{v}^{(0)} &=- (1-\phi)\pd{}{t}\left(\bm\nabla_X\cdot\mathbf{u}_g^{(1)}\right)- \phi\bm\nabla_X\cdot\mathbf{v}^{(1)},\label{eq:pe-mass-1}\\
 \phi\left(\mathbf{v}^{(0)}-\pd{\mathbf{u}_g^{(0)}}{t} \right)&=-\bm\nabla_xp^{(0)}-\bm\nabla_Xp^{(1)},\label{eq:pe-darcy-1}\\
 \bm\nabla_x\cdot\bm{\sigma}_g^{(0)}+ \bm\nabla_X\cdot\bm{\sigma}_g^{(1)}&=\bm0,\label{eq:pe-mtm-1}\\
 \bm{\sigma}_g^{(0)}&=- p^{(0)}\mathbf{I}+\mathbb{D}:\left(\bm\nabla_x\mathbf{u}_g^{(0)}+\bm\nabla_X\mathbf{u}_g^{(1)}\right).\label{eq:lin-elastic-D1-1}
 \end{align}
In the fibre region $\Omega_f$, at $\mathcal{O}(\varepsilon)$ equations \eqref{eq:fibre-mtm}--\eqref{eq:lin-elastic-C1} supply
\begin{align}
 \bm\nabla_x\cdot\bm{\sigma}_f^{(0)}+ \bm\nabla_X\cdot\bm{\sigma}_f^{(1)}&=\bm0,\label{eq:fibre-mtm-1}\\ \bm{\sigma}_f^{(0)}&=\mathbb{C}:\left(\bm\nabla_x\mathbf{u}_f^{(0)}+\bm\nabla_X\mathbf{u}_f^{(1)}\right),\label{eq:lin-elastic-C1-1}
\end{align}
while on $\partial\Omega_f$ the boundary conditions \eqref{eq:cont-stress-bc}--\eqref{eq:kbc} supply at $\mathcal{O}(\varepsilon)$
\begin{align}
 \bm\sigma_g^{(1)}\cdot\mathbf{n}&= \bm\sigma_f^{(1)}\cdot\mathbf{n},\label{eq:cont-stress-bc-1}\\
 \mathbf{u}_g^{(1)} &=\mathbf{u}_f^{(1)},\label{eq:cont-disp-bc-1}\\
 \left(\mathbf{v}^{(1)}-\pd{\mathbf{u}_g^{(1)}}{t}\right)\cdot\mathbf{n}&=0.\label{eq:kbc-1}
\end{align}
Physically speaking, both equations \eqref{eq:lin-elastic-D1-0} and \eqref{eq:lin-elastic-C1-0} represent a stress-free deformation on the microscale at leading order, which implies that $\mathbf{u}_f\oz$ and $\mathbf{u}_g\oz$ are rigid body transformations. The requirement that $\mathbf{u}_f\oz$ and $\mathbf{u}_g\oz$ are periodic in $\mathbf{X}$ further implies that this transformation cannot be a rotation. The deformation must, therefore, be a translation and so $\mathbf{u}_f\oz(\mathbf{x},t)$ and $\mathbf{u}_g\oz(\mathbf{x},t)$ are independent of $\mathbf{X}$.
Continuity of displacement on the cell-scale interface $\partial\hat{\Omega}_f$ at leading order \eqref{eq:cont-disp-bc-0} then implies that $\mathbf{u}_g\oz(\mathbf{x},t)=\mathbf{u}_f\oz(\mathbf{x},t)$.
Similarly, as noted above, equation \eqref{eq:pe-darcy-0} implies that $p\oz$ is independent of $\mathbf{X}$.

\section{Definition of cell problems}\label{sec:cell-problem}
Having established that the leading order displacements $\mathbf{u}_f\oz$ and $\mathbf{u}_g\oz$ are independent of the microscale, we now obtain the equations that govern the microscale variation at $\mathcal{O}(\varepsilon)$ in the displacements.
Periodicity enables us to understand the microscale behaviour by considering a single repeating cell. We identify the restriction of $\Omega_f$ to the single repeating cell by $\hat{\Omega}_f$ and likewise $\hat{\Omega}_g$ is the restriction of $\Omega_g$ to the single repeating cell.
To be clear, $\partial\hat{\Omega}_f$ identifies the interface between $\Omega_f$ and $\Omega_g$ found within a single repeating cell. An example of this cell geometry is shown in Figure \ref{fig:scaffold}(d).

Substituting \eqref{eq:lin-elastic-D1-1} into \eqref{eq:pe-mtm-0} and \eqref{eq:lin-elastic-C1-1} into \eqref{eq:fibre-mtm-0}, and recalling that the leading order displacements and pressure are independent of $\mathbf{X}$, we obtain
\begin{align}
\bm\nabla_X\cdot\left(\mathbb{D}:\bm\nabla_X\mathbf{u}_g^{(1)}\right) &=\bm0, &\mathrm{in}\,\,\hat{\Omega}_g,\label{eq:pe-mtm-0-D}\\
\bm\nabla_X\cdot\left(\mathbb{C}:\bm\nabla_X\mathbf{u}_f^{(1)}\right) &=\bm0, &\mathrm{in}\,\,\hat{\Omega}_f,\label{eq:fibre-mtm-0-C}
\end{align}
subject to the continuity of stress and displacement conditions
given by equations \eqref{eq:cont-stress-bc-0} and \eqref{eq:cont-disp-bc-1} on the cell-scale interface $\partial\hat{\Omega}_f$
\begin{align}
\left(\mathbb{C}:\bm\nabla_X\mathbf{u}_f^{(1)}-\mathbb{D}:\bm\nabla_X\mathbf{u}_g^{(1)}\right)\cdot\mathbf{n} &= -p^{(0)}\mathbf{n}-\left(\mathbb{C}:\bm\nabla_x\mathbf{u}_f^{(0)}-\mathbb{D}:\bm\nabla_x\mathbf{u}_f^{(0)}\right)\cdot\mathbf{n},\label{eq:cont-stress-bc-1-DC}\\
 \mathbf{u}_f^{(1)} &=\mathbf{u}_g^{(1)}.\label{eq:cont-disp-bc-1a}
\end{align}
Boundary conditions on the surface of the repeating cell are provided by requiring $\mathbf{u}_f^{(1)}$ and $\mathbf{u}_g^{(1)}$ to be periodic, with one additional boundary condition required to remove the translational freedom which is later set by requiring that various components of the microscale solution have zero mean on the microscale.

We note that equations \eqref{eq:pe-mtm-0-D} and \eqref{eq:fibre-mtm-0-C} define linear homogeneous problems, subject only to linear forcing by the leading order displacement, $\mathbf{u}_f^{(0)}$, and the leading order pressure, $p^{(0)}$, via the Neumann boundary condition \eqref{eq:cont-stress-bc-1-DC}. Hence, their solutions are of the form
\begin{align}
\mathbf{u}_g^{(1)}&=\mathbf{r}(\mathbf{X})p^{(0)}+\mathcal{B}(\mathbf{X}):\bm\nabla_x\mathbf{u}_f^{(0)},\label{eq:u-gel-1}\\ \mathbf{u}_f^{(1)}&=\mathbf{q}(\mathbf{X})p^{(0)}+\mathcal{A}(\mathbf{X}):\bm\nabla_x\mathbf{u}_f^{(0)},\label{eq:u-fibre-1}
\end{align}
where $\mathbf{r}$ and $\mathbf{q}$ are vectors and $\mathcal{B}$ and $\mathcal{A}$ are rank 3 tensors. The solutions \eqref{eq:u-gel-1} and \eqref{eq:u-fibre-1} are substituted into \eqref{eq:pe-mtm-0-D} and \eqref{eq:fibre-mtm-0-C}, respectively, and it  follows from the linearity of \eqref{eq:pe-mtm-0-D}  and \eqref{eq:fibre-mtm-0-C} that
\begin{align}
(\lambda_g+\mu_g)\bm\nabla_X\left(\bm\nabla_X\cdot\mathbf{r}\right)+\mu_g\bm\nabla^2\mathbf{r}&=\bm0,&\textrm{in}\,\,\hat{\Omega}_g, \label{eq:pe-mtm-0-r}\\
(\lambda_f+\mu_f)\bm\nabla_X\left(\bm\nabla_X\cdot\mathbf{q}\right)+\mu_f\bm\nabla^2\mathbf{q}&=\bm0,&\textrm{in}\,\,\hat{\Omega}_f, \label{eq:pe-mtm-0-q}
\end{align}
where we have exploited the constitutive (linearly elastic) assumptions for $\mathbb{D}$ and $\mathbb{C}$, specified by equations \eqref{eq:lin-elastic-D2} and \eqref{eq:lin-elastic-C2}, respectively. On the interface between the component materials equations \eqref{eq:pe-mtm-0-r}--\eqref{eq:pe-mtm-0-q} for $\mathbf{r}$ and $\mathbf{q}$ are subject to the boundary conditions
\begin{align} \left(\mathbb{C}:\bm\nabla_X\mathbf{q}-\mathbb{D}:\bm\nabla_X\mathbf{r}\right)\cdot\mathbf{n}& = -\mathbf{n}, &\mathrm{on}\,\,\partial\hat{\Omega}_f,\label{eq:stress-bc-qr}\\
 \mathbf{q}&=\mathbf{r},  &\mathrm{on}\,\,\partial\hat{\Omega}_f.
\end{align}
We additionally require that $\mathbf{r}$ and $\mathbf{q}$ are periodic in $\mathbf{X}$, and that
\begin{equation}
 \iiint_{\hat{\Omega}_g}\mathbf{r} \,\mathrm{d}V + \iiint_{\hat{\Omega}_f}\mathbf{q} \,\mathrm{d}V=\mathbf{0},\label{eq:no-mean-rq}
\end{equation}
so that the solution has zero mean on the microscale.
We note that equations \eqref{eq:pe-mtm-0-r}--\eqref{eq:no-mean-rq} for $\mathbf{r}$ and $\mathbf{q}$ define a linear elasticity problem on the  repeating cell in which deformations in the gel region $\hat{\Omega}_g$ and the fibre region $\hat{\Omega}_f$ are coupled and caused by a jump in stress at the interface between $\hat{\Omega}_g$ and $\hat{\Omega}_f$.

A similar procedure is applied to obtain governing equations for $\mathcal{B}$ and $\mathcal{A}$. We first rewrite the components of each rank 3 tensor in a vectorised form as
\begin{align}
\mathbf{b}^{(mn)}=\mathcal{B}_{imn}\mathbf{e}_i,\quad \textrm{and} \quad \mathbf{a}^{(mn)}=\mathcal{A}_{imn}\mathbf{e}_i,
\end{align}
where $\mathbf{e}_i$ are the Cartesian basis vectors, $m$, $n=1,2,3$, and we sum over the repeated index $i$. Substituting these vectorised forms into \eqref{eq:pe-mtm-0-D} and \eqref{eq:fibre-mtm-0-C}, and exploiting the linearity of these problems, we obtain
\begin{align}
(\lambda_g+\mu_g)\bm\nabla_X\left(\bm\nabla_X\cdot\mathbf{b}^{(mn)}\right)+\mu_g\bm\nabla^2\mathbf{b}^{(mn)}&=\bm0,&\textrm{in}\,\,\hat{\Omega}_g,\label{eq:pe-mtm-0-b}\\
(\lambda_f+\mu_f)\bm\nabla_X\left(\bm\nabla_X\cdot\mathbf{a}^{(mn)}\right)+\mu_f\bm\nabla^2\mathbf{a}^{(mn)}&=\bm0,&\textrm{in}\,\,\hat{\Omega}_f,
\end{align}
where we have again made use of the constitutive assumptions \eqref{eq:lin-elastic-D2} and \eqref{eq:lin-elastic-C2}. On the interface between the component materials, these problems for $\mathbf{b}^{(mn)}$ and $\mathbf{a}^{(mn)}$ are subject to the boundary conditions
\begin{align}
\left(\mathbb{C}:\bm\nabla_X\mathbf{a}^{(mn)}-\mathbb{D}:\bm\nabla_X\mathbf{b}^{(mn)}\right)\cdot\mathbf{n}&=-(\mathbb{C}:\mathbf{I}^{(mn)}-\mathbb{D}:\mathbf{I}^{(mn)})\cdot\mathbf{n}, &\mathrm{on}\,\,\partial\hat{\Omega}_f,\label{ABstressbc}\\
 \mathbf{b}^{(mn)}&=\mathbf{a}^{(mn)},  &\mathrm{on}\,\,\partial\hat{\Omega}_f,
\end{align}
where $\mathbf{I}^{(mn)}$ is an indicator matrix whose $(m,n)$-th entry is $1$, otherwise zero. We additionally require that  $\mathbf{b}^{(mn)}$ and $\mathbf{a}^{(mn)}$ are periodic in $\mathbf{X}$, and that
\begin{equation}
\iiint_{\hat{\Omega}_g}\mathbf{b}^{(mn)} \,\mathrm{d}V + \iiint_{\hat{\Omega}_f}\mathbf{a}^{(mn)} \,\mathrm{d}V=\mathbf{0},\label{eq:no-mean-ba}
\end{equation}
so that the microscale solution has zero mean. 
Thus, equations \eqref{eq:pe-mtm-0-b}--\eqref{eq:no-mean-ba} represent a further nine linear elasticity problems on the repeating cell in which deformations in the gel region $\hat{\Omega}_g$ and the fibre region $\hat{\Omega}_f$ are coupled, and caused by a jump in stress at the interface between $\hat{\Omega}_g$ and $\hat{\Omega}_f$.
 
A similar procedure is applied to determine  $p^{(1)}$, the $\mathcal{O}(\varepsilon)$ pressure of the fluid phase in the hydrogel region. We note that as $\mathbf{u}_f\oz=\mathbf{u}_g\oz$ is independent of $\mathbf{X}$, equation \eqref{eq:pe-mass-0} implies that the divergence of the fluid phase velocity in the poroelastic region is zero at leading order. We then take the divergence of \eqref{eq:pe-darcy-1} on the microscale to find that
\begin{equation}
\nabla_X^2p^{(1)}=0, \qquad\qquad \mathrm{in}\,\,\hat{\Omega}_g.\label{eq:p1-laplace}
\end{equation}
Next we take the scalar product of \eqref{eq:pe-darcy-1} with $\mathbf{n}$ and, exploiting equations \eqref{eq:cont-disp-bc-1} and \eqref{eq:kbc-1}, obtain the following boundary condition for $p^{(1)}$ on the hydrogel-fibre interface,
\begin{equation}
 \bm\nabla_Xp^{(1)}\cdot\mathbf{n} = -\bm\nabla_xp^{(0)}\cdot\mathbf{n}, \qquad\qquad  \mathrm{on}\,\,\partial\hat{\Omega}_f.\label{eq:p1-neumann}
\end{equation}
Thus, equations \eqref{eq:p1-laplace}--\eqref{eq:p1-neumann} comprise a linear homogeneous cell problem for $p^{(1)}$ subject to forcing by the leading order pressure $p^{(0)}$ via the Neumann boundary condition. As above, we formulate a solution to this problem as
\begin{equation}
 p^{(1)} =\mathbf{f}\cdot\bm\nabla_xp^{(0)},\label{eq:p1-sol}
\end{equation}
where $\mathbf{f}=\mathbf{f}(\mathbf{X})$ is a vector. Upon substitution of \eqref{eq:p1-sol} into \eqref{eq:p1-laplace} we obtain
\begin{equation}
 \nabla^2_X\mathbf{f}=0, \qquad \qquad \mathrm{in}\,\,\hat{\Omega}_g.\label{eq:f-laplace}
\end{equation}
Similarly, substitution of \eqref{eq:p1-sol} into \eqref{eq:p1-neumann} provides the boundary condition
\begin{equation}
\bm\nabla_X\mathbf{f}\cdot\mathbf{n} = -\mathbf{n}, \qquad \qquad \mathrm{on}\,\,\partial\hat{\Omega}_f.\label{eq:f-stress-bc}
\end{equation}
Finally, we require that $\mathbf{f}$ is periodic in $\mathbf{x}$, and that
\begin{equation}
 \iiint_{\hat{\Omega}_g}\mathbf{f} \,\mathrm{d}V=\mathbf{0},\label{eq:f-zero-mean}
\end{equation}
so that the microscale solution has zero mean. Thus, equations \eqref{eq:f-laplace}--\eqref{eq:f-zero-mean} define linear, scalar problems for the three components of $\mathbf{f}$.
 
\section{Macroscale equations and effective parameters}\label{sec:macroscale}

To complete the homogenisation procedure we now average across the microscale solutions from Section \ref{sec:cell-problem} to obtain governing equations and effective material parameters for the composite material at the macroscale. 

We integrate the $\mathcal{O}(\varepsilon)$ continuity of mass equation \eqref{eq:pe-mass-1} over the microscale repeating unit cell, and divide by the cell volume. It follows from the divergence theorem, and application of the continuity of displacement condition \eqref{eq:cont-disp-bc-1} and the kinematic condition \eqref{eq:kbc-1} that
\begin{equation}
 \phi\bm\nabla_x\cdot\mathbf{v}\eff + (1-\phi)\frac{|\hat{\Omega}_g|}{|\hat{\Omega}|}\pd{}{t}\left(\bm\nabla_x\cdot\mathbf{u}_f^{(0)}\right) =
\frac{1}{|\hat{\Omega}|}\pd{}{t}\iiint_{\hat{\Omega}_f}\bm\nabla_X\cdot\mathbf{u}_f^{(1)}\mathrm{d}V,\label{eq:mass-eff}
\end{equation}
where $|\hat\Omega|$ is the volume of a microscale repeating unit cell, $|\hat\Omega_g|$ is the volume within this cell occupied by the hydrogel and $\mathbf{v}\eff$ is the effective velocity of the fluid phase of the hydrogel, namely
\begin{equation}
 \mathbf{v}\eff(\mathbf{x},t) = \frac{1}{|\hat{\Omega}|}\iiint_{\hat{\Omega}_g}\mathbf{v}\oz(\mathbf{x},\mathbf{X},t) \,\mathrm{d}V.
\end{equation}
 We now substitute the solution for $\mathbf{u}_f\oo$ given by \eqref{eq:u-fibre-1} into the averaged continuity of mass equation \eqref{eq:mass-eff} to obtain 
\begin{equation}
 \phi\bm\nabla_x\cdot\mathbf{v}\eff + (1-\phi)\frac{|\hat{\Omega}_g|}{|\hat{\Omega}|}\pd{}{t}\left(\bm\nabla_x\cdot\mathbf{u}_f^{(0)}\right)
=\mathbf{S}\eff:\pd{}{t}\bm\nabla_x\mathbf{u}_f^{(0)} + \Gamma\eff\pd{p\oz}{t},\label{eq:mass-eff-1}
\end{equation}
where $\mathbf{S}\eff$ is an effective compressibility tensor (rank 2) and $\Gamma\eff$ is a parameter related to the compressibility of the composite material; \hl{this accounts for both the compressibility of the linear elastic materials in the composite (namely the PCL fibres and the solid phase of the hydrogel) as well as the effect associated with the flow of the incompressible fluid phase within the hydrogel due to the deformation of the solid phase (where water will be lost from the composite).}
These are defined as
\begin{align}
 \mathbf{S}\eff &= \frac{1}{|\hat{\Omega}|}\iiint_{\hat{\Omega}_f}\bm\nabla_X\cdot \mathcal{A}\,\mathrm{d}V,\label{eq:Seff}\\
 \Gamma\eff &= \frac{1}{|\hat{\Omega}|}\iiint_{\hat{\Omega}_f}\bm\nabla_X\cdot\mathbf{q} \,\mathrm{d}V.\label{eq:Gamma-eff}
\end{align}
To determine these effective parameters we first solve equations \eqref{eq:pe-mtm-0-r}--\eqref{eq:no-mean-rq} and \eqref{eq:pe-mtm-0-b}--\eqref{eq:no-mean-ba} to obtain $\mathcal{A}$ and $\mathbf{q}$ for a particular geometry and then use these solutions in \eqref{eq:Seff} and \eqref{eq:Gamma-eff} above.

Continuing, we integrate the $\mathcal{O}(\varepsilon)$ version of Darcy's law \eqref{eq:pe-darcy-1} over the microscale repeating cell, and divide by total cell volume to obtain
\begin{equation}
\phi\left(\mathbf{v}\eff - \frac{|\hat{\Omega}_g|}{|\hat{\Omega}|}\pd{\mathbf{u}_f^{(0)}}{t} \right) =-\frac{|\hat{\Omega}_g|}{|\hat{\Omega}|}\bm\nabla_xp^{(0)} - \frac{1}{|\hat{\Omega}|}\iiint_{\hat{\Omega}_g} \bm\nabla_Xp^{(1)}\mathrm{d}V.\label{eq:darcy-eff}
\end{equation}
We then use equation \eqref{eq:p1-sol} to substitute for $p^{(1)}$  in equation \eqref{eq:darcy-eff}. Rewriting the left-hand side of that equation in a more compact form, we obtain 
\begin{equation}
 \phi\left(\mathbf{v}\eff - \frac{|\hat{\Omega}_g|}{|\hat{\Omega}|}\pd{\mathbf{u}_f^{(0)}}{t}\right)  = -\mathbf{K}\eff\bm\nabla_xp^{(0)}, \label{eq:darcy-eff-1}
\end{equation}
where $\mathbf{K}\eff$ is an effective permeability tensor (rank 2) for the composite material; this is defined as
\begin{equation}
 \mathbf{K}\eff = \frac{1}{|\hat{\Omega}|}\left(|\hat{\Omega}_g|\mathbf{I} + \iiint_{\hat{\Omega}_g} \bm\nabla_X\mathbf{f}\,\mathrm{d}V\right).\label{eq:Keff}
\end{equation}
Thus, to determine the effective permeability $\mathbf{K}\eff$ we first solve \eqref{eq:f-laplace}--\eqref{eq:f-zero-mean} to obtain $\mathbf{f}$ for a particular microscale geometry and then use that solution in \eqref{eq:Keff}. In later numerical simulations it is convenient to eliminate $\mathbf{v}^{\textrm{eff}}$ by substituting \eqref{eq:darcy-eff-1} into \eqref{eq:mass-eff-1} to give
\begin{align}
-\mathbf{K}\eff\nabla_x^2p^{(0)}  + \frac{|\hat{\Omega}_g|}{|\hat{\Omega}|}\pd{}{t}\left(\bm\nabla_x\cdot\mathbf{u}_f^{(0)}\right)
=\mathbf{S}\eff:\pd{}{t}\bm\nabla_x\mathbf{u}_f^{(0)} + \Gamma\eff\pd{p\oz}{t}.\label{eq:mass-eff-2}
\end{align} 
We remark  that writing the equation in this form eliminates the porosity $\phi$, obviating the need to know that quantity.

Finally, we integrate the $\mathcal{O}(\varepsilon)$ conservation of momentum equations, \eqref{eq:pe-mtm-1} and \eqref{eq:fibre-mtm-1}, over the microscale repeating unit cell and divide by the total cell volume; we then apply continuity of stress at the hydrogel-fibre interface \eqref{eq:cont-stress-bc-1} to obtain a volume averaged conservation of momentum equation
\begin{align}
 \bm\nabla_x\cdot\bm\sigma\eff =\frac{1}{|\hat{\Omega}|}\left(\bm\nabla_x\cdot\iiint_{\hat{\Omega}_f}\bm\sigma_f\oz\,\mathrm{d}V + \bm\nabla_x\cdot\iiint_{\hat{\Omega}_g}\bm
\sigma_g\oz\,\mathrm{d}V\right) =\bm0,\label{eq:mtm-eff}
\end{align}
where $\bm\sigma^{\textrm{eff}}$ is an effective stress tensor (rank 2) representing the macroscale stress of the composite material. To develop an explicit expression for $\bm\sigma^{\textrm{eff}}$ we substitute the first order displacements, \eqref{eq:u-gel-1} and \eqref{eq:u-fibre-1}, into the definitions of leading-order stress, \eqref{eq:lin-elastic-D1-1} and \eqref{eq:lin-elastic-C1-1}, to obtain
\begin{align}
\bm\sigma_g^{(0)}&=-p^{(0)}\mathbf{I}+\mathbb{D}:\left(\bm\nabla_x\mathbf{u}_f^{(0)}+\left(\bm\nabla_X\mathbf{r}\right)p^{(0)}+\left(\bm\nabla_X\mathcal{B}\right):\bm\nabla_x\mathbf{u}_f^{(0)}\right),\\
\bm\sigma_f^{(0)}&=\mathbb{C}:\left(\bm\nabla_x\mathbf{u}_f^{(0)}+\left(\bm\nabla_X\mathbf{q}\right)p^{(0)}+\left(\bm\nabla_X\mathcal{A}\right):\bm\nabla_x\mathbf{u}_f^{(0)}\right).
\end{align}
On substituting these expressions into \eqref{eq:mtm-eff} we deduce that the appropriate form of the effective stress tensor is
\begin{align}
 \bm\sigma\eff &= \mathbb{C}\eff:\bm\nabla_x\mathbf{u}_f^{(0)} + \mathbf{G}\eff p^{(0)},\label{eq:stress-eff}
\end{align}
where $\mathbb{C}\eff$ is an effective elasticity tensor (rank 4), and $\mathbf{G}^{\textrm{eff}}$ is a rank 2 tensor describing the hydrostatic component of the effective stress; these are defined as
\begin{align}
\mathbb{C}\eff & = \frac{1}{|\hat{\Omega}|}\left(|\hat{\Omega}_f|\mathbb{C}+|\hat{\Omega}_g|\mathbb{D}+ \mathbb{C}:\iiint_{\hat{\Omega}_f}\bm\nabla_X\mathcal{A}\,\mathrm{d}V+\mathbb{D}:\iiint_{\hat{\Omega}_g}\bm\nabla_X\mathcal{B}\,\mathrm{d}V\right),\ \label{eq:Ceff}\\
\mathbf{G}\eff& = \frac{1}{|\hat{\Omega}|}\left( - |\hat{\Omega}_g|\mathbf{I}+ \mathbb{C}:\iiint_{\hat{\Omega}_f}\bm\nabla_X\mathbf{q}\,\mathrm{d}V + \mathbb{D}:\iiint_{\hat{\Omega}_g}\bm\nabla_X\mathbf{r}\,\mathrm{d}V\right),\label{eq:Geff}
\end{align}
where $|\hat\Omega_f|$ is the volume occupied by the fibres. Thus, to find the effective stress tensor $\bm\sigma\eff$ of the macroscale composite material for a particular (microscale) hydrogel-fibre geometry we first solve equations \eqref{eq:pe-mtm-0-r}--\eqref{eq:no-mean-rq} and \eqref{eq:pe-mtm-0-b}--\eqref{eq:no-mean-ba} to obtain the solution components of the microscale cell problem, namely  $\mathbf{r}$, $\mathbf{q}$, $\mathcal{B}$ and $\mathcal{A}$, and then use these solutions in expressions \eqref{eq:Ceff} and \eqref{eq:Geff} above.

To summarise, we have now derived a system of four macroscale equations for continuity of mass \eqref{eq:mass-eff-1}, Darcy's law \eqref{eq:darcy-eff-1}, conservation of momentum \eqref{eq:mtm-eff} and the effective stress tensor \eqref{eq:stress-eff} which govern the macroscale variables for displacement $\mathbf{u}_f^{(0)}$, pressure in the hydrogel $p^{(0)}$ and the effective velocity of the fluid phase of the hydrogel $\mathbf{v}\eff$.




\subsection{Simplifications due to cell symmetry and linear elasticity}\label{sec:siimplication}


Many entries in the tensors defining the macroscale properties derived in the previous section can be shown to vanish either by arguments due to the symmetry of the cell geometry, and/or by exploiting our assumptions that the fibres and the solid phase of hydrogel are linearly elastic. 

The domain of the microscale repeating cell is $0\leqslant X\leqslant 1$, $0\leqslant Y\leqslant 1$, $0\leqslant Z\leqslant \theta$, where $\theta=h/l$ is the dimensionless microscale height of the cell.
Within this cell the fibres are arranged so that there are two half cylinders, with non-dimensional radius $\rho=d/(2l)$, with mid-lines along ($Y=0.5$, $Z=0$) and ($Y=0.5$, $Z=\theta$) respectively. There is a cylinder, with non-dimensional radius $\rho$, with its mid-line along ($X=0.5$, $Z=\theta/2$).
The union of the cylinder and the two half cylinders form the elastic fibre region $\hat{\Omega}_f$.
The complement of $\hat{\Omega}_f$ in the repeating box is the hydrogel region $\hat{\Omega}_g$. A measure of the overlap between the fibres is then $\theta/(4\rho)$, where this quantity is equal to 1 if adjacent fibre layers are just touching, and equal to $1/2$ if they completely overlap. 
A schematic diagram of the cell geometry showing both the hydrogel and fibre regions is shown in Figure~\ref{fig:scaffold}(e).



The arrangement of the fibres is such that the cell geometry is symmetric about all three mid-planes and it is therefore only necessary to consider an eighth of the cell volume. The symmetries in the components of the surface normal vector $\mathbf{n}=n_i\mathbf{e}_i$ are as follows
\begin{align}
n_1,&\quad \textrm{odd in }X, \textrm{ even in }Y, Z,\\
n_2,&\quad \textrm{odd in }Y, \textrm{ even in }X, Z,\\
n_3,&\quad \textrm{odd in }Z, \textrm{ even in }X, Y.
\end{align}
When combined with the microscale boundary conditions \eqref{eq:stress-bc-qr}, \eqref{ABstressbc} and  \eqref{eq:f-stress-bc}, these impose further symmetries on the solution components $\mathbf{q}$, $\mathbf{r}$, $\mathcal{A}$, $\mathcal{B}$ and $\mathbf{f}$. For example, consider the $X$-component of the boundary condition \eqref{eq:f-stress-bc}
\begin{align}
\pd{f_1}{X}n_1+\pd{f_1}{Y}n_2+\pd{f_1}{Z}n_3 = -n_1,\qquad \qquad \mathrm{on}\,\,\partial\hat{\Omega}_f,
\end{align}
where each product on the left hand side must be odd in $X$, and even in $Y$ and $Z$ to match the normal component $n_1$ on the right hand side. This implies that
\begin{align}
\pd{f_1}{X},&\quad \textrm{even in }X, Y, Z,\label{eq:df1dX}\\
\pd{f_1}{Y},&\quad \textrm{odd in } X, Y, \textrm{ even in } Z,\\
\pd{f_1}{Z},&\quad \textrm{odd in } X, Z, \textrm{ even in } Y.
\end{align}
Therefore, when each of these quantities is integrated over the hydrogel (cell) volume in \eqref{eq:Keff} to calculate $\mathbf{K}^{\text{eff}}$ only the integral of the $X$-derivative \eqref{eq:df1dX} above is non-zero since it is even in all three dimensions.
Applying a similar argument to the other components of $\mathbf{f}$, we find that in \eqref{eq:f-stress-bc} only three of the the nine components of the volume integral terms are non-zero (listed in Table~\ref{table:derivs-etc}), and that $\mathbf{K}^{\text{eff}}$ is diagonal.
\begin{table}
\begin{tabular}{ccc}
variable & derivatives even in all dimensions & no. non-zero (total) \\
\hline\hline  $\mathbf{f}$ & $\displaystyle\pd{f_1}{X}$, $\displaystyle\pd{f_2}{Y}$, $\displaystyle\pd{f_3}{Z}$ & 3 (9) \\
\hline$\mathbf{q}$ & $\displaystyle\pd{q_1}{X}$, $\displaystyle\pd{q_2}{Y}$, $\displaystyle\pd{q_3}{Z}$ & 3 (9) \\
\hline$\mathbf{r}$ & $\displaystyle\pd{r_1}{X}$, $\displaystyle\pd{r_2}{Y}$, $\displaystyle\pd{r_3}{Z}$ & 3 (9) \\
\hline \centering$\mathbf{a}^{(mn)}$ & \shortstack{$\displaystyle\pd{a_1^{(mm)}}{X}$,$\displaystyle\pd{a_2^{(mm)}}{Y}$,$\displaystyle\pd{a_3^{(mm)}}{Z}$ (for $m=1,2,3$),\\ $\displaystyle\pd{a_1^{(12)}}{Y}$, $\displaystyle\pd{a_2^{(12)}}{X}$, $\displaystyle\pd{a_1^{(13)}}{Z}$, $\displaystyle\pd{a_1^{(13)}}{X}$, $\displaystyle\pd{a_2^{(23)}}{Z}$, $\displaystyle\pd{a_3^{(23)}}{Y}$} & 15 (81)\\
\hline$\mathbf{b}^{(mn)}$ & as for $\mathbf{a}^{(mn)}$ & 15 (81)\\
\hline
\end{tabular}
\caption{List of derivatives of the components of the microscale solution from Section \ref{sec:cell-problem} which are even in all spatial dimensions, and so make a non-zero contribution to the volume integrals used to calculate the effective parameters in  Section \ref{sec:macroscale}.}\label{table:derivs-etc}
\end{table}

An identical argument is applied to the boundary conditions \eqref{eq:stress-bc-qr} for $\mathbf{q}$ and $\mathbf{r}$. Although the form of this boundary condition is slightly more complicated, the symmetry properties of the components are the same as for $\mathbf{f}$, so calculation of $\mathbf{G}^{\text{eff}}$ in \eqref{eq:Geff} only requires the evaluation of six volume integrals (listed in Table~\ref{table:derivs-etc}), and we note here that $\mathbf{G}^{\text{eff}}$ is diagonal.

A similar procedure is performed for $\mathbf{a}^{(mn)}$ and $\mathbf{b}^{(mn)}$, although it is necessary to first exploit the symmetry properties of $\mathbb{C}$ and $\mathbb{D}$. The constitutive assumptions that both the fibres and the solid phase of the hydrogel are linearly elastic, \eqref{eq:lin-elastic-D2} and \eqref{eq:lin-elastic-C2}, mean that $\mathbb{C}$ and $\mathbb{D}$ are right-symmetric and left-symmetric (that is $\mathbb{C}_{ijkl}=\mathbb{C}_{ijlk}$ and $\mathbb{C}_{ijkl}=\mathbb{C}_{jikl}$). It follows from the right-symmetry of $\mathbb{C}$ and $\mathbb{D}$, and from \eqref{ABstressbc}, that 
\begin{alignat}{5}
\mathbf{a}^{(12)}&=\mathbf{a}^{(21)},\quad &\mathbf{a}^{(13)}&=\mathbf{a}^{(31)}, \quad &\mathbf{a}^{(23)}&=\mathbf{a}^{(32)}, \\
\mathbf{b}^{(12)}&=\mathbf{b}^{(21)} ,\quad&\mathbf{b}^{(13)}&=\mathbf{b}^{(31)}, \quad&\mathbf{b}^{(23)}&=\mathbf{b}^{(32)},
\end{alignat}
which reduces the number of required calculations; it is only necessary to consider 18 components of the rank 3 tensors $\mathcal{A}$ and $\mathcal{B}$, rather than the full 27. 

The left-symmetry of $\mathbb{C}$ and $\mathbb{D}$ implies that the right-hand side of each component of \eqref{ABstressbc} is proportional to a single component of the normal vector $\mathbf{n}$. Careful consideration of each component of this boundary condition reveals the symmetry properties of the various derivatives of $\mathbf{a}^{(mn)}$ and $\mathbf{b}^{(mn)}$ which are required for calculation of $\mathbb{C}^{\text{eff}}$ in \eqref{eq:Ceff}. Only 15 of the original 81 volume integrals are non-zero and these are listed in Table~\ref{table:derivs-etc}. To summarise, by exploiting the geometric cell symmetry and linearly elastic constitutive relations only 39 of the original 189 volume integrals need to be evaluated to calculate the effective macroscale properties.

\section{Solution procedure and comparison with experiments}\label{sec:results-etc}

We now validate the model presented in Sections~\ref{sec:cell-problem} and \ref{sec:macroscale} against a series of experiments that were performed to establish how fibre spacing affects the mechanical properties of reinforced hydrogel scaffolds. These experiments involved scaffolds reinforced with PCL fibres 20\,$\mu$m in diameter and 3D-printed at spacings of either 300 or 800\,$\mu$m (with three replicates for each choice of fibre spacing). The fibre lattices are then cast in GelMA to produce cylindrical scaffolds with diameters between 5.54--5.98\,mm and heights 1.80--1.98\,mm. These composite samples were held in unconfined compression at a fixed strain between two parallel plates while the applied stress required to maintain this displacement was recorded; after an initial ramping phase the required stress decreases slowly due to the poroelastic relaxation of the composite.

Details of the numerical solution procedure for the microscale cell problem of Section~\ref{sec:cell-problem} and the homogenised macroscale problem of Section~\ref{sec:macroscale} are given in Sections \ref{sec:result-micro} and \ref{sec:results-macro}, respectively. The experimental relaxation tests are compared to our theoretical simulation results in Sections \ref{sec:relaxtion-test}, with a focus on replicating the poroelastic relaxation phase of these experiments in the simulations.

\subsection{Microscale solution procedure}\label{sec:result-micro}

\begin{table}
  \tabcolsep=8pt
  \begin{tabular}{cccc}
  \hline\hline
    Dimensional fibre spacing & $\rho$ & $\theta$  & $\theta/4\rho$\\
  \hline
    $300\mu$m  & $0.0333$  & $0.1066$ &  $0.8$\\
    $800\mu$m  & $0.0125$  & $0.04$   &  $0.8$\\
  \hline\hline
  \end{tabular}
\caption{Dimensionless parameters characterising the repeating cell for the \hl{two} repeating cell geometries used in the simulations. The fibre radius is $\rho$, the dimensionless microscale height of the cell is  $\theta=h/l$, and $\theta/4\rho$  is a measure of the vertical overlap between adjacent fibre layers, where a larger value indicates less overlap and $\theta/(4\rho)=1$ represents the case where the fibres are just touching.}\label{tab:cellgeom}
\end{table}

The microscale cell problem requires the solution of the linear elasticity problems \eqref{eq:pe-mtm-0-r}--\eqref{eq:no-mean-rq} and \eqref{eq:pe-mtm-0-b}--\eqref{eq:no-mean-ba} to obtain $\mathbf{r}$, $\mathbf{q}$, $\mathbf{b}^{(mn)}$ and $\mathbf{a}^{(mn)}$, and the solution to Laplace's equation (and boundary conditions) \eqref{eq:f-laplace}--\eqref{eq:f-zero-mean} to obtain $\mathbf{f}$. These sub-problems are solved using the multi-physics package COMSOL, which can perform computations on the interpenetrating geometry of the fibres and hydrogel regions. These computations are repeated for two cell geometries, described by the dimensionless parameter values given in Table \ref{tab:cellgeom}. 

For each of the cell geometries in Table~\ref{tab:cellgeom} we use the COMSOL simulation results to calculate $\mathbb{C}\eff$, $\mathbf{G}$, $\mathbf{K}\eff$, $\mathbf{S}\eff$ and $\Gamma\eff$. As described in Section~\ref{sec:siimplication}, this requires the computation of only the volume integrals of the derivatives of the solution components given in Table~\ref{table:derivs-etc}.
The form of the effective elasticity tensor $\mathbb{C}\eff$ in \eqref{eq:Ceff} reveals that the composite material can best be described as an orthotropic material in which two of the defined directions $X$ and $Y$ of the effective material properties of the composite are the same.
This is not the same as a transversely isotropic material which has one distinguishable axis and is isotropic in any plane which lies perpendicular to that axis.
In our material $X$ and $Y$ are interchangeable, but the two directions which are parallel to the directions of the fibres are both `special' directions.
This is intuitively simple to reconcile with the square grid pattern of the printed fibres.
From these calculations we observe that $\mathbb{C}\eff_{1111}=\mathbb{C}\eff_{2222}$ are an order of magnitude larger than $\mathbb{C}\eff_{3333}$, indicating that the composite material is much stronger along the fibre directions than perpendicular to the fibres.
The other nonzero components of $\mathbb{C}\eff$ are much smaller, which suggests that the composite material would be weaker in shearing.

\subsection{Macroscale solution procedure}\label{sec:results-macro}

Having obtained the effective material parameters from the microscale problem, we proceed to solve the macroscale equations \eqref{eq:mass-eff-1}, \eqref{eq:darcy-eff-1}, \eqref{eq:mtm-eff} and \eqref{eq:stress-eff} with a finite element scheme. We aim to compare this with experiments on a cylindrical scaffold and the dimensions of the scaffolds from these experiments determines the choice of length scale $L$. For example, for the experiments with 300\,$\mu$m fibre spacing we take this length scale to be $L=5.76$\,mm, the mean diameter of the three scaffolds, and the corresponding mean dimensionless scaffold height is $\eta=H/L=0.34$. The solution domain is then
\begin{equation}
 x^2+y^2 \leqslant (1/2)^2, \qquad\qquad 0\leqslant z\leqslant \eta.\label{macrodomain}
\end{equation}
The scaffold is held between two plates, so no-slip conditions are appropriate at both the upper and lower surfaces of the cylinder. Additionally, we prescribe a time-dependent displacement in $z$ on the upper surface as a means of implementing the loading strategy. The appropriate boundary conditions are then
\begin{align}
& \mathbf{u}_f^{(0)}=\mathbf{0}, \qquad &\pd{p^{(0)}}{z}=0, \qquad &\mathrm{on}\,\, z=0,\label{eq:macro-noslip}\\
& u_{f1}^{(0)}=u_{f2}^{(0)}=0, \qquad u_{f3}^{(0)}=d(t), \quad &\pd{p^{(0)}}{z}=0, \qquad &\mathrm{on}\,\, z=\eta.
\end{align}
Different choices of the displacement function $d(t)$ are required to simulate the relaxation tests, and these will be defined in the following section. We impose no stress boundary conditions on the curved surfaces of the cylinder
\begin{equation}
 p^{(0)}=0, \qquad \boldsymbol{\sigma}\eff\cdot\mathbf{e}_R =0 \qquad \mathrm{on}\,\, x^2+y^2=(1/2)^2,\label{macBCp}
\end{equation}
where $\mathbf{e}_R$ is the outward-pointing unit normal to the cylinder surface.

We calculate numerical solutions of  \eqref{eq:mass-eff-1}, \eqref{eq:darcy-eff-1}, \eqref{eq:mtm-eff} and \eqref{eq:stress-eff}, subject to the boundary conditions given by Equations \eqref{eq:macro-noslip}--\eqref{macBCp} using a finite element method.
The domain $x^{2}+y^{2} \leq (1/2)^2$, $0\leq z\leq \eta$ is partitioned into tetrahedral elements using the mesh generation package tetgen \cite{tetgen}. We then eliminate $\mathbf{v}\eff$ from this system of equations by substituting an explicit relation for $\mathbf{v}\eff$ obtained from substitution of \eqref{eq:darcy-eff} into \eqref{eq:mass-eff-1}. 
A finite element solution is then calculated, using an implicit approximation to all time derivatives, that uses a quadratic approximation to $\mathbf{u}_f^{(0)}$ on each element, and a linear approximation to $p^{(0)}$ on each element.  This finite element method has been shown to be stable for poroelasticity \cite{Murad}, and is therefore suitable here since the homogenised governing equations are of a similar form to those that describe small deformation poroelasticity.

\subsection{Comparison with relaxation test experiments}\label{sec:relaxtion-test}

\begin{figure}
\includegraphics[width=0.75\textwidth]{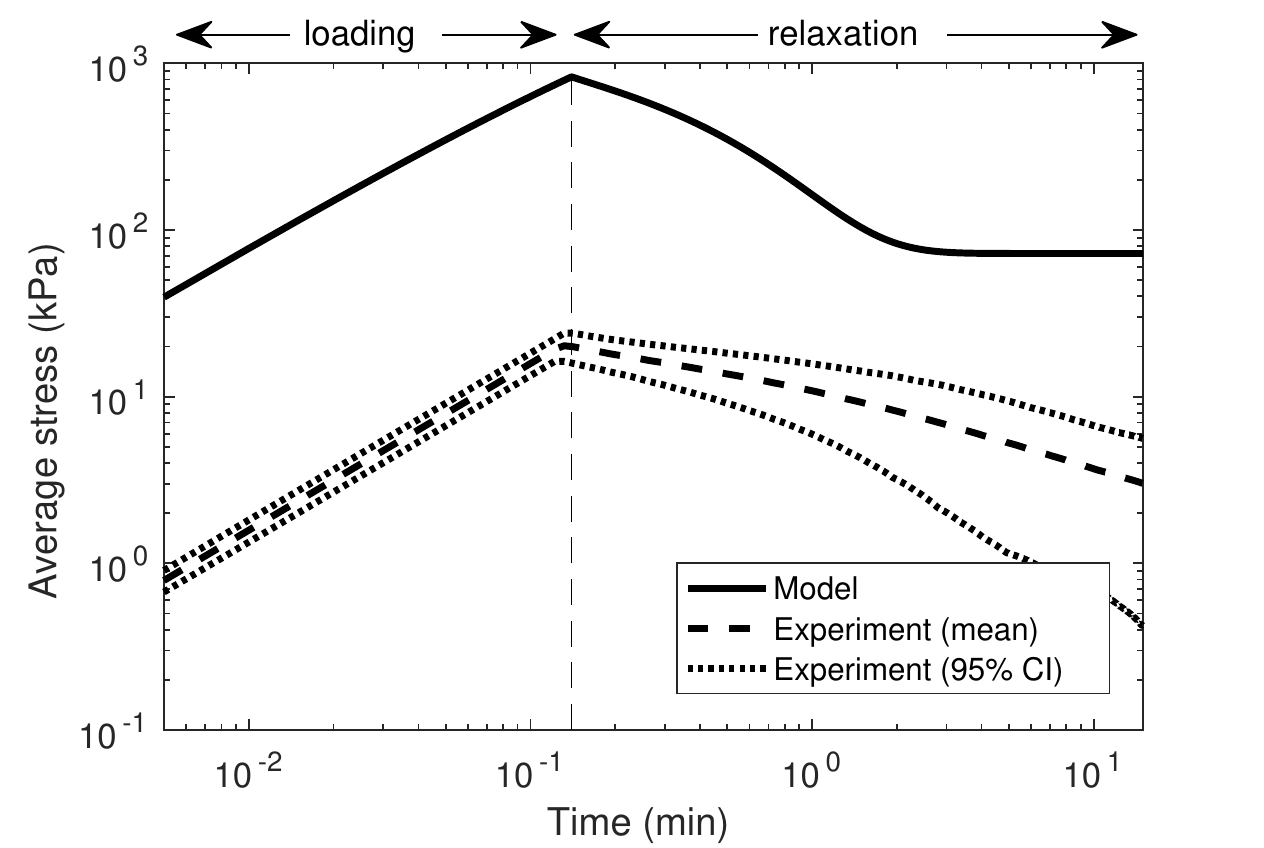}
\caption{Numerical simulations of the relaxation text for 300\,$\mu$m fibre spacing held at 6\% strain (solid line), shown as the time-dependent stress response of the scaffold to the imposed displacement given in \eqref{eq:relax-displacement}. Also shown is the mean time-dependent stress from three replicates of the experimental relaxation test (dashed line) and a 95\% confidence interval on this data (dotted lines).}\label{fig:relax-num}
\end{figure}

\begin{figure}
(a)\\\includegraphics[width=0.75\textwidth]{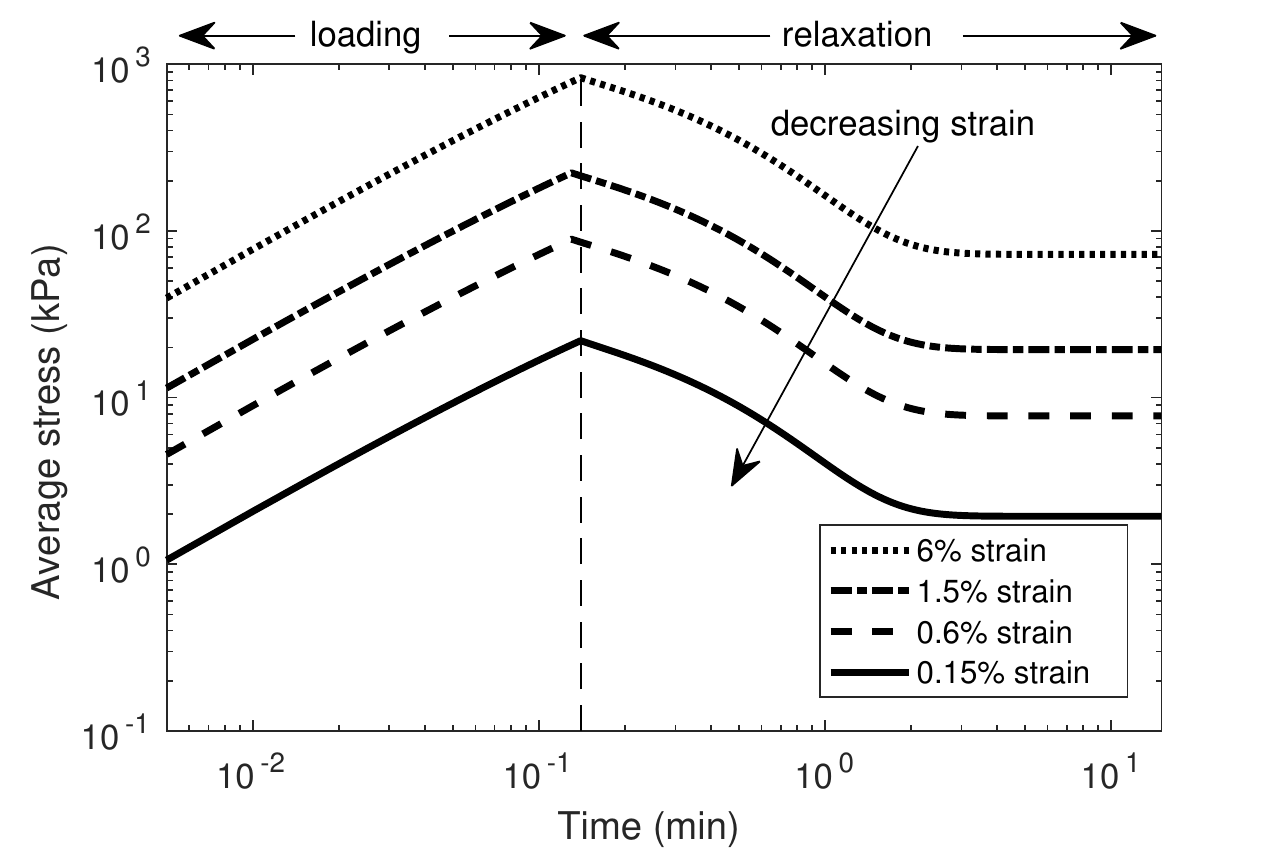}\\
(b)\\\includegraphics[width=0.75\textwidth]{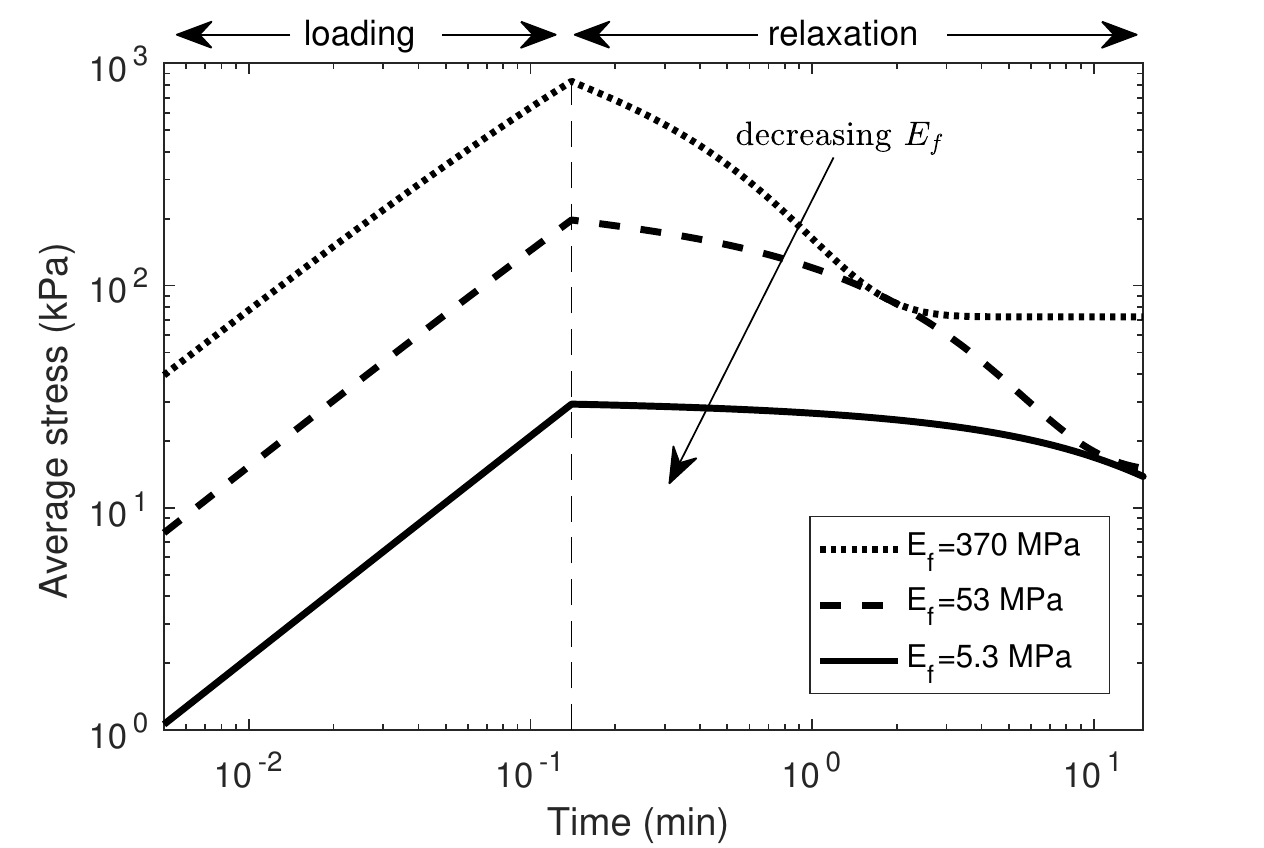}\\
(c)\\\includegraphics[width=0.75\textwidth]{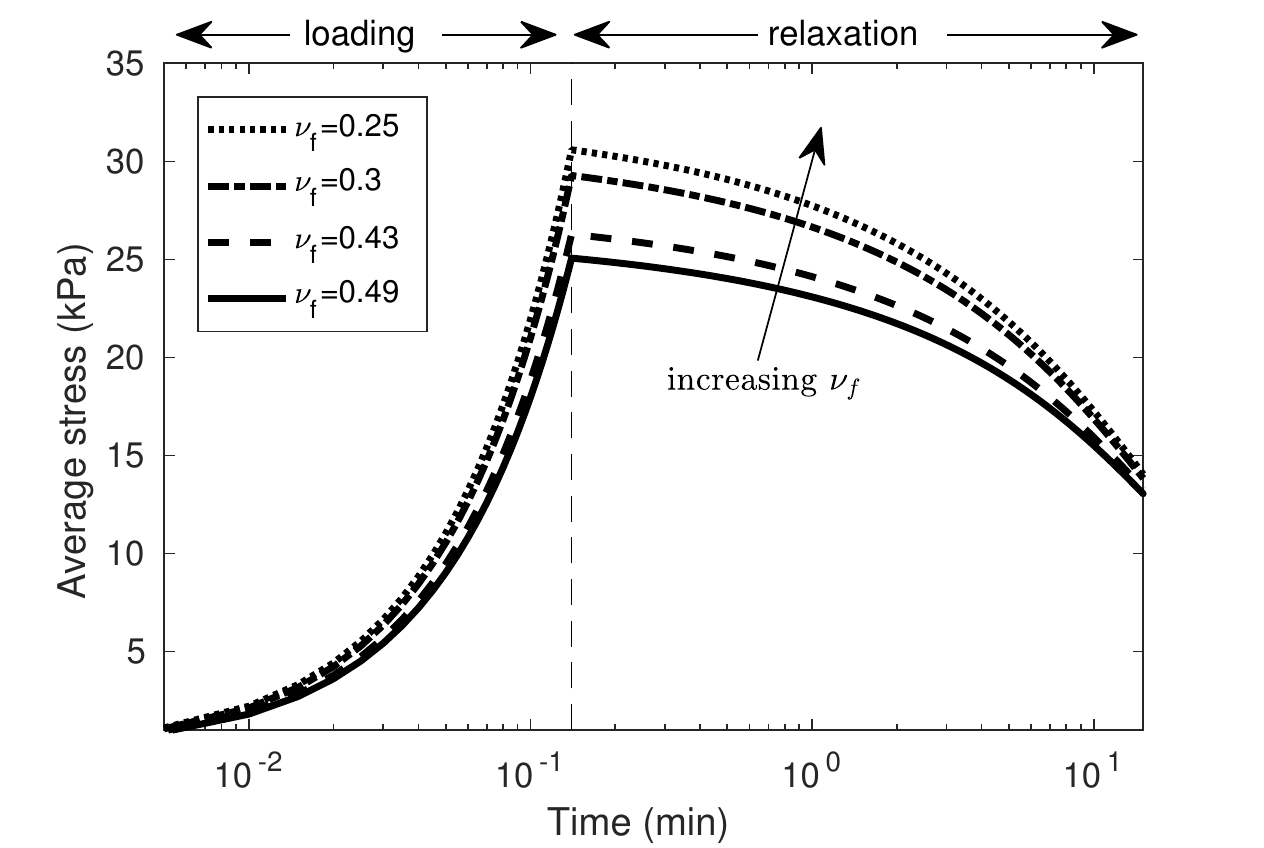}
\caption{Examples of the sensitivity of the stress response to the key parameters of the model. (a) Varying the displacement $\epsilon$, with the fibre parameters fixed at $E_f=363.3$\,MPa and $\nu_f=0.43$. (b) Varying the Young's modulus of the fibres $E_f$, with an applied strain of 6\% and $\nu_f=0.43$. (c) Varying the Poisson's ratio of the fibres $\nu_f$, with an applied strain of 6\% and $E_f=2.65$\,MPa.
}\label{fig:relax-sensitivity}
\end{figure}

\begin{figure}
(a)\\\includegraphics[width=0.75\textwidth]{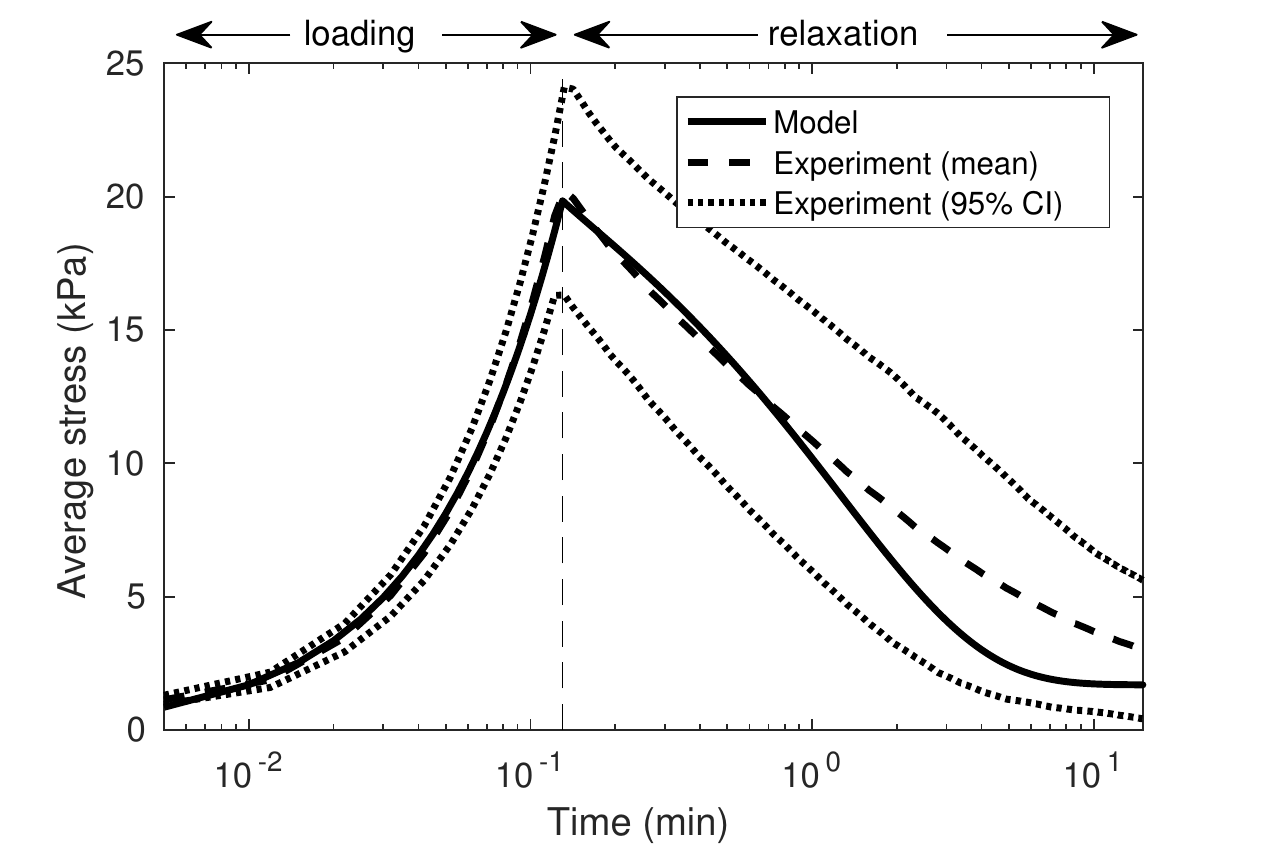}\\
(b)\\\includegraphics[width=0.75\textwidth]{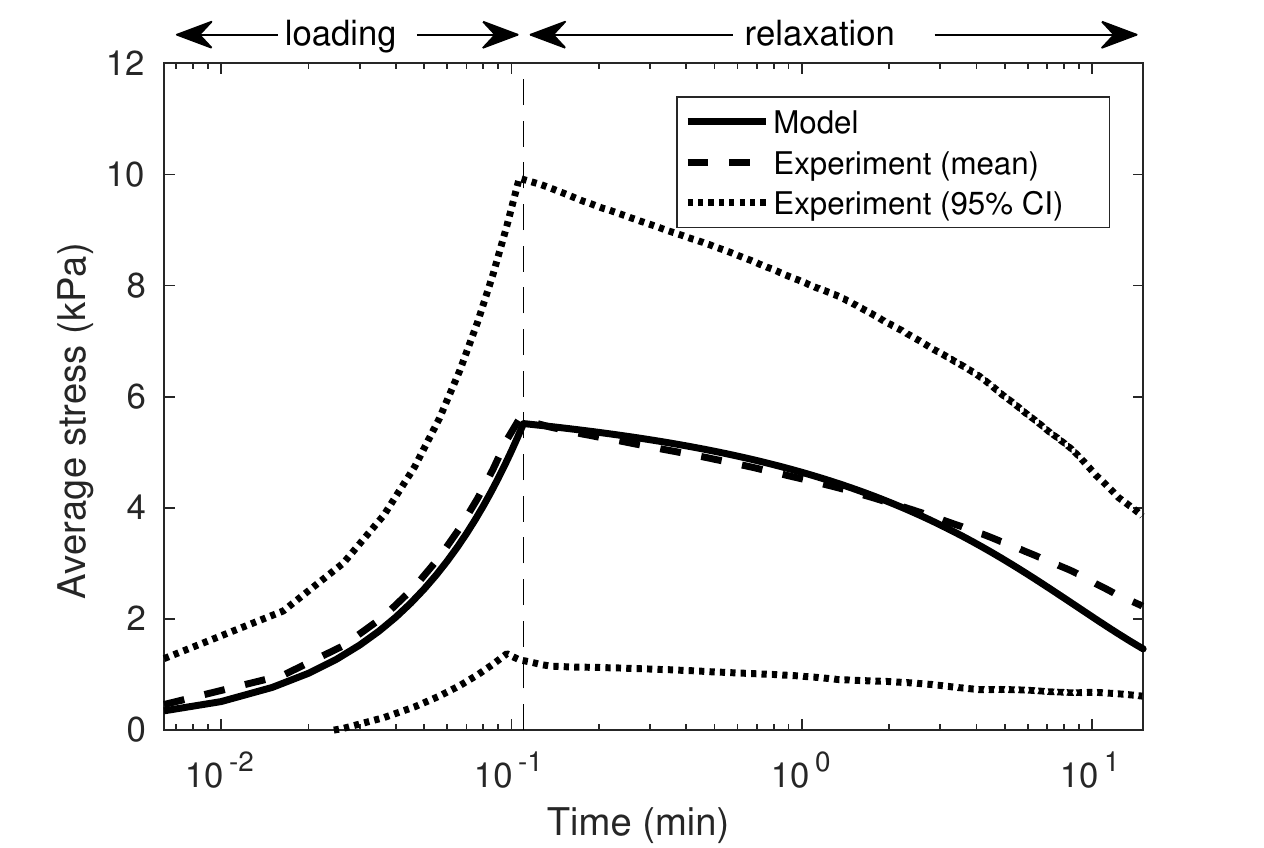}
\caption{Numerical simulations of the relaxation text compared with the experimental data. (a) 300\,$\mu$m  fibre spacing with adjusted modelling parameters of $\epsilon=0.45$\%, $E_f=90.8$\,MPa and $\nu_f=0.49$ (solid line). (b) 800\,$\mu$m fibre spacing with adjusted modelling parameters of $\epsilon=0.525$\%, $E_f=45.4$\,MPa and $\nu_f=0.49$ (solid line). Both (a) and (b) show the mean time-dependent stress from three replications of the experimental relaxation test (dashed line) and a 95\% confidence interval on this data (dotted lines). Note that 300\,$\mu$m fibre spacing data is the same as that shown in Figure \ref{fig:relax-num} with a log scale on the vertical axis.}.\label{fig:exp-comp-good}
\end{figure}


The relaxation test involves applying a 6\% strain at the top of the scaffold, and recording the stress required to maintain this displacement over the course of 15\,mins, that is for $0\leq t\leq15$.  In line with the experiments, the time-dependent displacement $d(t)$ of the top loading plate used in the simulations was chosen so that a strain of 6\% was attained after an initial period of linear displacement over $0\leq t<\delta$, where $\delta$ is a short initiation time. The form of the loading function for the relaxation test is then
\begin{align}
d(t)&=\epsilon\eta\left(\frac{1}{\delta}t H(\delta-t)+H(t-\delta)\right),\label{eq:relax-displacement}
\end{align}
where $\epsilon=0.06$, $H(t)$ is the Heaviside step function, and $\delta$ takes a slightly different value for each choice of fibre spacing to match the initial transient strain applied in the experiments;  these values are $\delta=0.14$ for the 300\,$\mu$m fibre spacing  and $\delta=0.11$ for the 800\,$\mu$m fibre spacing. 

The results of the macroscale simulations of this relaxation test for 300\,$\mu$m spacing are shown in  Figure~\ref{fig:relax-num}, along with experimental data based on three replicates of this test. The model exhibits qualitatively similar behaviour to the experiments, with an initial ramp up phase followed by a relaxation phase. These results are all displayed in terms of `average stress', defined as the total force applied at the top of the scaffold divided by the cross-sectional area. During the initial fast loading phase the response of the scaffold is dominated by the fibres, and so the average stress is essentially linearly elastic. During the relaxation phase the scaffold exhibits poroelastic behaviour due to the flow induced in the fluid phase of the hydrogel. There are marked quantitative differences between the experiments and simulations. The model overestimates the maximum stress attained after the initial loading by two orders of magnitude, and displays a more rapid relaxation, reaching a steady state after approximately 2 minutes, whereas the measured experimental stress is still decreasing at 15 minutes.

Close inspection of the scaffolds used in the experiments suggests some possible explanations for these discrepancies. The printed fibre lattices do not exactly correspond to our idealised model, with the fibres in the uppermost layers sagging and adopting a curved shape, as shown in Figures \ref{fig:scaffold}(b) and \ref{fig:scaffold}(c). We hypothesise that when the scaffold is loaded these fibres do not come under tension as readily as the fibres in the lower layers, and propose to account for this by adjusting the Young's modulus $E_f$ of all the fibres in the model. We also note that the Poisson's ratio of PCL fibres $\nu_f$ is not well characterised in the literature, with various sources assuming that this parameter falls in the range $0.3<\nu_f<0.49$ (see \cite{Castilho2017a,Eschbach1994,Eshraghi2010}).

Additionally, as a result of casting the printed fibres in the hydrogel, there is a thin layer of pure (unreinforced) hydrogel at the top of the scaffold. We hypothesise that this thin layer will yield more readily to loading than the reinforced gel below it, and that the reinforced gel, therefore, experiences a lower strain than that applied to the scaffold as a whole. For instance, if the depth of the pure hydrogel layer is 5--10\% of the height of the entire construct, then the strain applied to the reinforced part of the hydrogel may be less than 1\%. We propose to account for this by adjusting the applied strain in the model, via the parameter $\epsilon$ in \eqref{eq:relax-displacement}.

We now consider the effect of varying the three parameters described above, namely $\epsilon$, $E_f$, and $\nu_f$, on the time dependent average stress predicted by the model. The role of changing the applied strain is shown in Figure~\ref{fig:relax-sensitivity}(a), for four values of $\epsilon$ ranging from the recorded value of 6\% to a much smaller strain of 0.15\%. The peak stress value at the end of the loading phase for the smallest of these applied strains is approximately two orders of magnitude smaller than the original 6\% strain, and of the same order of magnitude as the experimentally recorded stress. The shape of the relaxation profile is, however, uneffected by varying the strain; it still decays more rapidly than the experimentally observed profile. 

The effect of lowering the Young's modulus of the fibres $E_f$ is shown in Figure~\ref{fig:relax-sensitivity}(b) for three choices of this parameter, with the original 6\% strain. To achieve a peak stress which is similar in magnitude to the experimentally observed value, $E_f$ must be set to a value which is an order or magnitude smaller than the lowest published value of this parameter. At the lowest published value (of $E_f=53$\,MPa) the model overpredicts the peak value of stress, but the relaxation profile is similar to that seen experimentally. The effect of varying $\nu_f$ is shown in Figure~\ref{fig:relax-sensitivity}(c); this has a relatively small effect on both the peak stress at the end of the loading phase and the rate at which the composite relaxes.

The sensitivity of the stress response to $\epsilon$ and $E_f$ shown in Figures~\ref{fig:relax-sensitivity}(a) and (b), suggests that the model will come close to the observed stress if both parameters are lowered in combination. We have performed a sparse parameter sweep through these parameters to determine values which produce reasonable agreement with the observed data. The stress given by these parameters is shown in Figure~\ref{fig:exp-comp-good}(a). Here, the parameters $\epsilon=$0.45\%, $E_f=90.8$\,MPa and $\nu_f=0.49$, produce a stress through the loading phase which closely follows the experiment, and a relaxation phase which is in good agreement up to time of about $t=1$, after which the model predicts a faster decay in stress.

Further relaxation tests were performed for reinforced hydrogel scaffolds with fibres printed at a wider spacing of 800\,$\mu$m, and data from three replicates of this experiment are compared to the homogenised model in Figure~\ref{fig:exp-comp-good}(b).  Applying the model naively as in Figure \ref{fig:relax-num} over-predicts the observed average stress by two orders of magnitude. The model solution shown in Figure \ref{fig:exp-comp-good}(b) is for $\epsilon=$0.525\%, $E_f=45.4$\,MPa and $\nu_f=0.49$; these parameters were obtained through a sparse parameter sweep, as described earlier. As before, the model follows the observations closely through the loading phase, and remains in agreement with the stress in the relaxation phase for a longer time than in the 300\,$\mu$m case. This agreement was obtained using a value of $E_f$ which was half that of the 300\,$\mu$m case, suggesting that the sagging of the fibres is more pronounced for this larger fibre spacing.

\section{Discussion}
\label{sec:discuss}

We have used mathematical homogenisation theory to develop a new model to describe the deformation of a composite elastic-poroelastic material. This was motivated by a desire to determine the macroscale mechanical properties of fibre-reinforced hydrogels used in the tissue engineering of articular cartilage. Our model enables us to calculate the effective material properties of the composite given knowledge of the material parameters of the constituent materials (namely the GelMA hydrogel and the PCL fibres) and the geometrical arrangement of the fibres and hydrogel within a single repeating cell of the composite. Our initial application of the model, shown in Figure~\ref{fig:relax-num}, predicted much stronger fibre-reinforced composites than those we tested experimentally, but exhibited good qualitative agreement with both the initial linear elastic loading phase and a poroelastic relaxation phase. Further numerical solutions, shown in Figure~\ref{fig:relax-sensitivity}, demonstrate that the predicted stiffness of the composite is very sensitive the Young's modulus of the PCL fibres and the strain applied to the composite. 


There are several possible explanations for why our model over-predicts the strength of the fibre-reinforced composites. In Section~\ref{sec:relaxtion-test} we discussed two of these in detail, namely the sagging of the printed fibres, which effectively lowers the Young's modulus of the fibres, and the presence of a layer of unreinforced hydrogel at the top of the scaffold, which means that an applied strain is not directly passed on to the reinforced composite material. Accounting for these effects, we obtained good agreement between the observed relaxation behaviour of the composite and our model, as shown in Figure~\ref{fig:exp-comp-good}. We postpone formally including these effects in the model for future work. Adding the extra thin layer of hydrogel would be a relatively straightforward extension of the current model. Accounting for the effect of sagging fibres would be more involved; in this case the cell geometry is no longer symmetric in $z$ and therefore some of the computational advantages that this symmetry confers would be lost.

There are other possible sources of discrepancy between the model and experiments.
For instance, the vertical spacing between the fibres was estimated with knowledge of the total number of printed layers and the overall scaffold height. If the vertical overlap between the fibres in the definition of the cell geometry is further reduced then the model predictions may be brought closer to the experimental data. Another possible source of the discrepancy is the boundary conditions imposed between the fibres and the hydrogel. We have assumed continuity of stress and displacement at the interface between the fibres and the hydrogel. In practice, a boundary condition that allows for some slip between the hydrogel and the fibres may be more appropriate, and would probably lead to the model predicting a weaker fibre-reinforced composite. Modifying the homogenisation procedure to account for such effects is an interesting direction for future work.

Finally, the hydrogel may not be perfectly poroelastic. Some of the observed relaxation behaviour may be due to viscous relaxation and incorporating these effects by using a different model for the hydrogel would require altering the homogenisation process. Such a model would introduce history dependence of the material and could potentially make it far less numerically efficient if a new set of cell problems had to be solved at each time step; see, for example, the discussion in \cite{Penta2014}.

The elastic material in our composite is much stronger than the poroelastic hydrogel; $\mu_f$ and $\lambda_f$ are five orders of magnitude larger than $\mu_g$ and $\lambda_g$, which might suggest that it is possible to  neglect entirely the contribution of the poroelastic region and model only the elastic fibre scaffold. This approach would not, however, capture the time dependent response of the composite. Accounting for both the elastic and the dynamic poroelastic nature of these composites, as we do here, is important to understand their mechanical properties. An interesting direction for study might be to formally incorporate the difference in the material properties of the hydrogel and the fibres in the model by exploiting the small parameter associated with the ratio of the Young's moduli of the elastic phase of the hydrogel and that of the fibres, and then repeating the homogenisation procedure.

Our model captures the key features of the fibre-reinforced hydrogel, in particular its orthotropic nature, and directly relates the material properties of the constituent hydrogel and fibres to those of the composite material. Modelling the mechanical properties of these scaffolds is an important step to inform tissue engineers about the stress experienced by cells when the scaffold is mechanically loaded, thus allowing future modelling work to consider the response of the cells to this stimulation. A key point of interest here is to understand how the scaffold is remodelled as the seeded cells deposit extracellular matrix components in response to loading, a process which eventually leads to implants which resemble natural articular cartilage. This might involve replacing the hydrogel phase with a cartilage-like phase that can explicitly describe the mechanical role of the extracellular matrix components; see \mbox{\cite{Klika2016}} for a review of such models of cartilage. Candidate models for this replacement phase include the model of \mbox{\cite{Mow1980}}, which treats cartilage as a poroelastic material, or the detailed cartilage model of \mbox{\cite{Ateshian2007}}, which includes the mechanical effects of ions interacting with the extracellular matrix.
Since the cells embedded in the scaffold are actively remodelling their surrounding mechanical environment, this approach should also account for the growth of the cartilage, and a natural framework to do this would be via the theory of morphoelasticity \mbox{\cite{Goriely2007}}.

To conclude, this homogenised model successfully captures the orthotropic nature of the fibre-reinforced hydrogel scaffold, can (when suitably adjusted) predict the behaviour seen in experimental relaxation tests and provides a basis for future study of the mechanical stimulation of cell-loaded scaffolds.

\begin{acknowledgements}
The research leading to these results has received funding from the European Union Seventh Framework Programme (FP7/2007-2013) under grant agreement no.~309962  (HydroZONES). The authors gratefully thank the Utrecht-Eindhoven strategic alliance and the European Research Council (consolidator grant 3D-JOINT, no.~647426) for the financial support.
\end{acknowledgements}

\appendix

\section{Calculation of Lam\'{e} parameters for PCL and GelMA}\label{sec:AppendixA}
We assume that the PCL fibres are an isotropic linear elastic material, with Young's modulus $E'_f$ and Poisson's ratio $\nu_f$. The published values of Young's modulus for printed PCL fibres vary with the method of printing and radius of the fibre, and so we will assume that $E_f'$ is between 53\,MPa and 363\,MPa (for details see \cite{Baker2016,Visser2015,Tan2005,Castilho2017a}), and that the Poisson's ratio $\nu_f$ is between 0.3 and 0.49 (see \cite{Eschbach1994,Eshraghi2010,Castilho2017a}).
The dimensional Lam\'{e} parameters $\mu'_f$ and $\lambda'_f$ in Table \ref{table:params} are calculated from these values as follows
\begin{equation}
 \mu_f' = \frac{E_f'}{2(1+\nu_f)}, \qquad\qquad \lambda'_f = \frac{E_f'\nu_f}{(1+\nu_f)(1-2\nu_f)}.\label{lamedefn}
\end{equation}

\begin{figure}
        \includegraphics[width=0.75\textwidth]{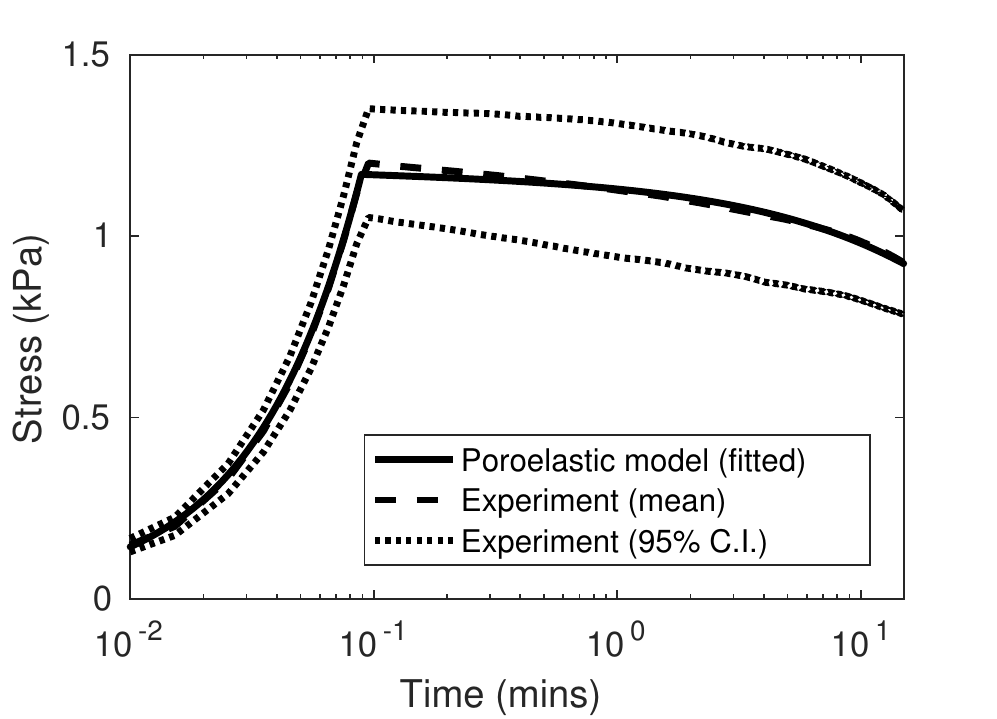}
    \caption{The time dependent stress response of three replications of an experimental relaxation test on unreinforced GelMA (mean as a black dashed line, 95\% confidence interval in black dotted lines) is shown against the numerical solution of the poroelastic equations for this relaxation test  with the fitted parameters of $E_g'=49.12$\,kPa, $\nu_g=0.23$ and $k'/\mu'=2.382\times10^{-4}$\,mm$^2$kPa$^{-1}$min$^{-1}$ (black solid line). In the experiments a cylinder of the hydrogel is held in unconfined compression between two parallel plates at 6\% displacement for 15\,mins.}\label{fig:gelMA-test}
\end{figure}

We have performed unconfined compression tests on GelMA to establish the Lam\'{e} parameters for the hydrogel. These tests were identical to the `relaxation' test performed on the reinforced composite (see Section \ref{sec:relaxtion-test}). Here, a pure GelMA cylinder of radius $2.5$\,mm and height $2$\,mm is held at $6\%$ strain between two parallel plates, with the stress required to maintain this displacement recorded over 15\,mins. Three replications of this test were performed and the results are shown in Figure~\ref{fig:gelMA-test}(a).

To obtain the Lam\'{e} parameters and the effective permeability $k'/\mu'$, this data is calibrated against finite element simulations of a poroelastic cylinder held at 6\% strain between parallel plates. Here, we solve \eqref{eq:pe-mass-dim}--\eqref{eq:lin-elastic-D2-dim} for a cylinder of hydrogel (with no reinforcement), assuming that there is no slip between the hydrogel and the plates, and thus obtain the average stress as a function of time. This simulation was repeated for a range of Poisson's ratio $\nu_g=0.2$--$0.3$ in intervals of 0.05 and a range of dimensionless Young's modulus $E_g=1\times10^{-3}$--$5\times10^{-3}$ in intervals of $10^{-4}$, where each choice of parameters has a characteristic relaxation profile. A fitted value of effective permeability $k'/\mu'$ is then used to dimensionalise these solutions to minimise the mean square error between each individual simulation and the three replications of the experiment. The combination of material parameters which minimise the mean squared error is (once dimensionalised) $E'_g=49.1$kPa, $\nu_g=0.23$ and $k'/\mu'=2.382\times10^{-4}$mm$^2$kPa$^{-1}$min$^{-1}$. The corresponding Lam\'{e} parameters $\mu'_g$ and $\lambda'_g$ are given in Table~\ref{table:params}. The simulated relaxation test for these parameters is shown as a dashed line in Figure~\ref{fig:gelMA-test}(a) and is in agreement with the experimental data which demonstrates that it is both reasonable and accurate to consider the hydrogel as a poroelastic material.

\label{lastpage}

\bibliographystyle{acm}
\bibliography{ReinforcedHydrogelHomogPaperR6}

\end{document}